\documentclass[aps,prd,groupedaddress,epsfig]{revtex4-1}
\usepackage{graphicx}% Include figure files
\usepackage{color}
\usepackage{amsfonts}
\newcommand{\cno}{$\mathsf{CNO}$}
\newcommand{\nesi}{$\mathsf{Ne}$-$\mathsf{Si}$}
\newcommand{\p}{$\mathsf{p}$}
\newcommand{\he}{$\mathsf{He}$}
\newcommand{\fe}{$\mathsf{Fe}$}
\begin{document}
\title{Universal cosmic rays energy spectrum and the mass composition at the 'ankle' and above}
\author{T. Wibig}
%\thanks{}
\affiliation{Faculty of Physics and Applied Informatics, University of Lodz, 90-256 {\L }\'{o}d\'{z}, Poland.}
\author{A.W. Wolfendale}
%\thanks{}
\affiliation{Department of Physics, Durham University, Durham, DH1 3LE, UK}

\date{\today}

\begin{abstract}
 We have used recently published data from the Pierre Auger Observatory and the Telescope Array Project to conclude some
inferences  concerning the origin and composition of ultra high-energy cosmic rays (UHECR). We advocate for model the UHECR 
flux as a combination of the Galactic and 
Extragalactic components exchanging the dominant role at the famous feature of the energy spectrum called 'the ankle'. We put a special emphasis on the individual experiment energy resolution 
and possible biases it introduces. We have eventually combined the data from both big experiments to the one 'world average' UHECR 
spectrum and we used it, supported by the 
analysis of the distribution of the Extensive Air Shower (EAS) development maximum depth to find 
the UHECR mass composition of both: Galactic and Extragalactic components. 
We studied the dependence of the on mass composition on the cosmic ray source distribution in space. 
We present results for some models existing in the literature.
We found the evidence of the deficit of Extragalactic sources in the vicinity of the Galaxy of the  the UHECR, 
what is in concordance with the small scale anisotropy analysis results.
\end{abstract}
\pacs{98.70.Sa, 96.50.sb, 96.50.sd, 98.65.-r}
% 95.85.Ry	Neutrino, muon, pion, and other elementary particles; cosmic rays
%98.70.Sa	Cosmic rays (including sources, origin, acceleration, and interactions)
%96.50.S-	Cosmic rays 
%96.50.sb	Composition, energy spectra and interactions
%96.50.sd	Extensive air showers
%98.65.-r	Galaxy groups, clusters, and superclusters; large scale structure of the Universe

\maketitle

\section{Introduction}
The question about the nature of the particles of energies
around the spectral feature called the `ankle'  is one of the most important
and still open questions in the Ultra High-Energy Cosmic
Ray (UHECR) domain. The lower energy cosmic rays are
known to be mainly confined within the Galaxy, and it is
known that the Galactic field is too weak to contain particles
of energies above $10^{19}$ eV. It is believed by us and others
that the `ankle' is related to the Extragalactic (EG) origin of the
cosmic ray particles of ultra high energies.

The two biggest experiments Pierre Auger Observatory (PAO) and Telescope Array Project (TA) acting on the UHECR domain at present have been publishing the new data successively, for some time. 
The statistics of collected events increased substantially the accuracy of delivered results. 
Actually both groups gathered numbers of events which allow them to obtain the CR spectra with the accuracy 
limited mainly not by the aperture, but rather by the methodology of the data analysis including the accuracy of the Monte 
Carlo simulations (see, e.g., \cite{1475-7516-2013-02-026}). Further increase of statistics would improve, in principle, the determination of some particular shower 
characteristics as,e.g., the average depth of the shower maximum ($X_{\rm max}$) or its dispersion (rms$_{X\rm max}$), but the general 
problems with theoretical description is still based and hung on simulations. The analysis of all existing data suggests that the problem 
now is not in the statistical uncertainty, but rather of the systematics in both: the measurements and the EAS 
modeling. 

\begin{figure}
 \includegraphics[width=8.5cm]{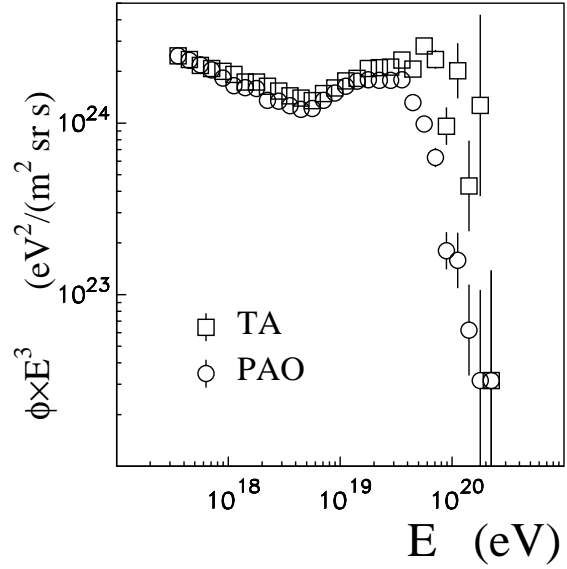}
\caption{Pierre Auger Observatory and Telescope Array Project UHECR energy spectra as they are reported recently by both groups.
\label{paota}}
\end{figure}

\begin{figure}
\centerline{
 \includegraphics[width=8.5cm]{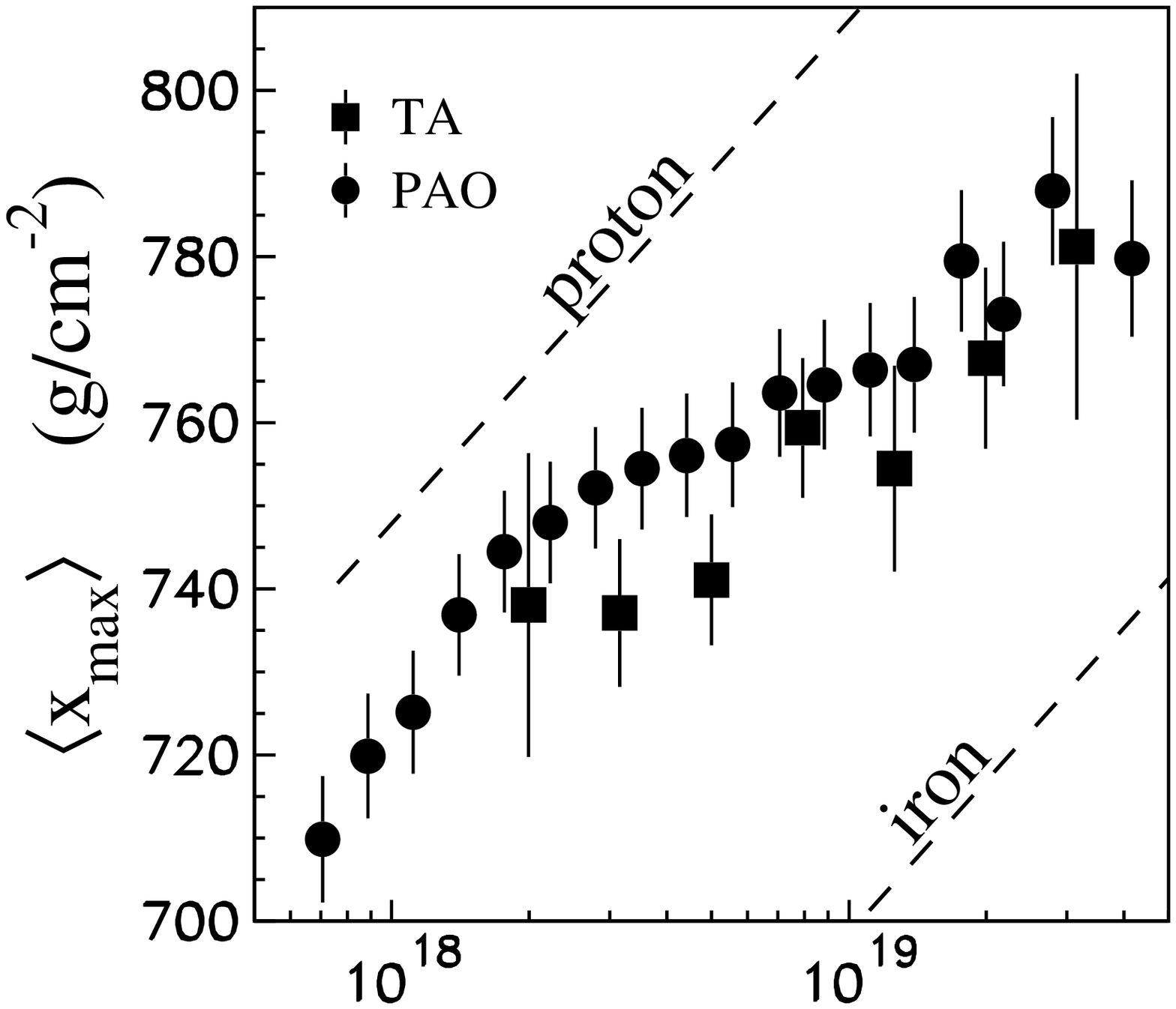}
 \includegraphics[width=8.5cm]{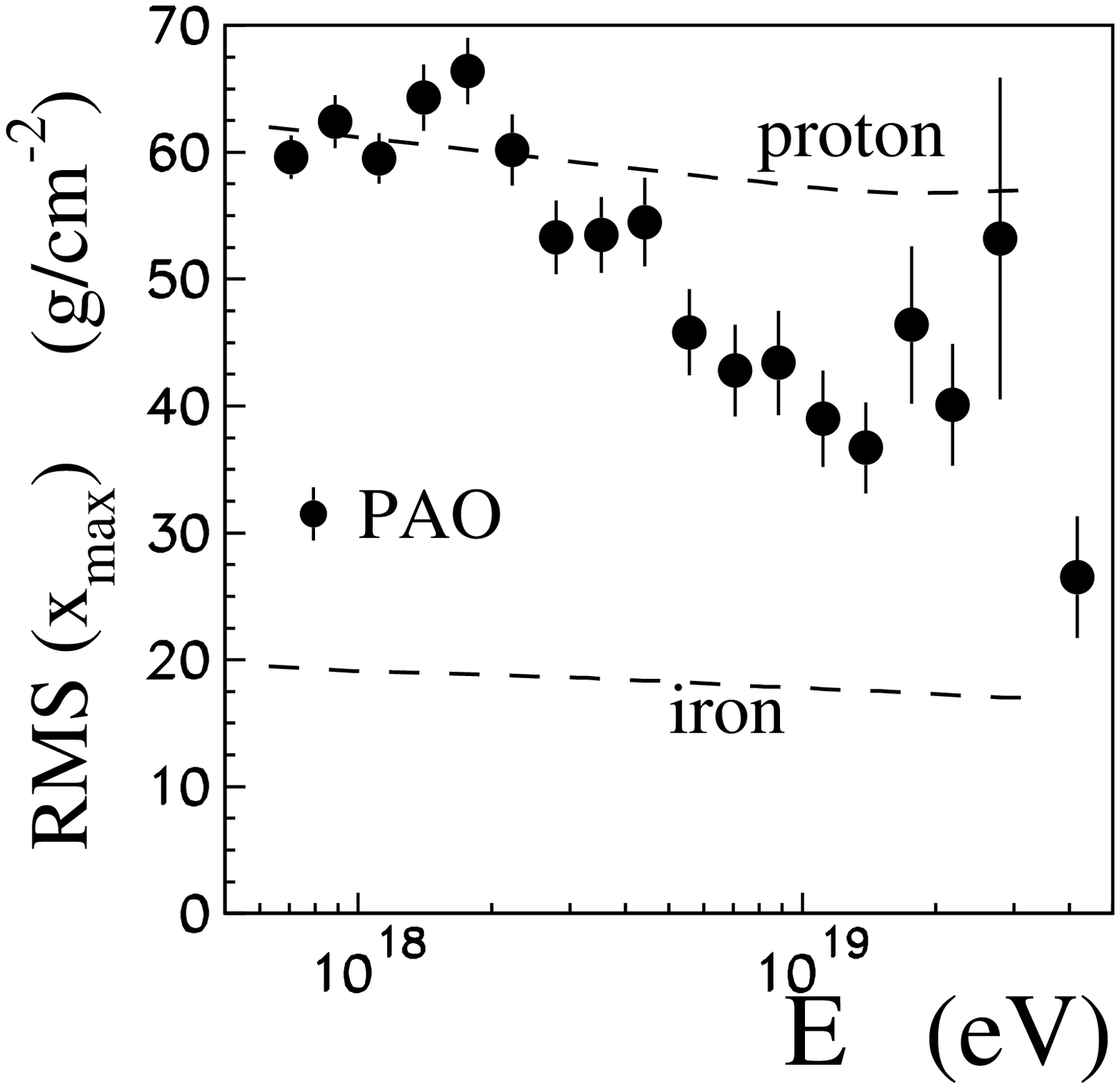}}
\caption{PAO and TA data of the distribution of depth of the UHECR  shower maxima: the average position of $X_{\rm xmax}$
and its spread: rms$_{X\rm max}$. Lines shows results of simulations for pure proton and iron showers made by the PAO Collaboration.\label{xmaxdata}}
\end{figure}

The recently published data from PAO \cite{PhysRevD.90.122006, paospectrum, Abraham2010239, Settimo2012} 
and TA \cite{taspesumm, Abbasi2016131, Abbasi201549, Tinyakov201429, 2041-8205-768-1-L1}  form the solid basis to search for conclusions 
concerning the origin and composition of UHECR (see, e.g., \cite{PhysRevD.90.122005}). 
We would like to present here 
our analysis of data on 
cosmic ray spectra (shown in Fig.\ref{paota}) taken together together. We combined them to the one 'world average' UHECR spectrum and used as a one much more accurate datum which supported by the
analysis of the $\langle X_{\rm max}\rangle$ and rms$_{X\rm max}$ data measured also by both experiments which we 
show in Fig.\ref{xmaxdata}. The recent analysis of the widths of the $X_{\rm max}$ distribution for narrow energy
bins use the whole distributions as they are measured. The TA group does not report rms$_{X\rm max}$ data. Published distributions are not corrected for the instrumental effects and their widths are considerably wider than values reported by other groups (e.g. HiRes \cite{PhysRevLett.104.161101} or PAO \cite{PhysRevD.90.122005}), so we will use below only PAO data on rms$_{X\rm max}$.

t is known for many years that spectra published by both big experiments
do not agree with each other. Some unknown experimental bias (at least in the estimated shower energy) still
exists and it is widely discusses, e.g., in Ref.~\cite{1475-7516-2017-04-038}. Based on the recent analysis of both teams the bias is of order of $\sim0.1$ in the $\log(E)$ scale what gives about $\sim 20$\% (14\% \cite{paospectrum}) of the energy over(under)estimation error. There is of course no indication which spectrum should be used as a datum. 
Therefore our analysis will be presented two fold: once assuming that the PAO data is the one, once the TA measurement.
We do not wish to judge which solution is closer to the reality, but we are going, first, to test what the uncertainty it makes
concerning the mass composition of the `ankle'.

 The conclusions of the recent PAO paper  \cite{1475-7516-2017-04-038} and also by Taylor, Ahlers and Hooper \cite{PhysRevD.92.063011, Hooper2010151},
   Globus, Allard  and Parizot  \cite{PhysRevD.92.021302} and others are obtained using only the PAO measurements.
    The TA Collaboration conclusions were formed, contrary, using only TA results \cite{Tinyakov201429, Fukushima:2015bza}.
 We would like to perform the examination of both: the spectra and composition data
 measured by 
 the Pierre Auger Observatory and the Telescope Array experiment in the whole 'ankle' region where, in our opinion, the Galactic and Extragalactic components meet.

 Our results and comparison with the other work conclusions will be commented in Sec.\ref{discusion}.

\section{The UHECR flux model}
d the way proposed and successfully applied in Ref. \cite{Szabelski2002125}.
The cosmic ray flux is assumed to be the sum of Galactic and Extragalactic components which cross at 
about the middle of the structure called `the ankle', located roughly between the particle energies of $10^{18}$ 
and 10$^{19}$ eV. Below the ankle, we assumed, that the cosmic rays originate mostly in our Galaxy. Galactic 
magnetic fields are strong enough and enough turbulent to confined cosmic ray 
particles  of energies below $10^{18}$ eV and propagate them in the diffusive regime within the volume of the Galactic halo.
Depending on the electric charge the confinement ends around $10^{18}$ eV and particles of higher energies 
leak out to intergalactic space. 
The Galactic (G) component energy spectrum is assumed to be a power-law with the index of 
about (-3) observed below 10$^{17}$ eV modified by the rigidity dependent Galactic leakage factor which we have found to be roughly 
of the Gaussian shape \cite{Szabelski2002125}. The real structure of the Galactic component cut-off is probably much more complicated. 
At least it should be a sum of different nuclei fluxes, each with the cut-off at different total energy due to the 
most probable rigidity dependent mechanism of the leakage. The contribution of the source acceleration energy limit(s) 
is also possible here. 
Due to the lack of knowledge of all this processes we assumed just pure Gaussian cut-off shape with two 
parameters left to be adjusted in the fitting procedure. The form of the Galactic component is given by Eq. (\ref{fg}).

\begin{equation}
                \Phi_{\rm G  }(E)=  %        {d  \Phi_{\rm G} \over d E}= 
                {{\Phi_0}_{\rm G}} \:E^{-\gamma_G}\: \times \:
\left\{
\begin{array}{ll}
1 
&
\rm if\ {E<E_{\rm Gcut}}                 \\
\exp \left( -\left( {E-E_{\rm Gcut}}\right)
                                                      ^2 \: /2  \sigma_{\rm G}^2\right)
&
\rm if\ {E>E_{\rm Gcut}}
\end{array}
\right.
\label{fg}
\end{equation}
The Extragalactic (EG) component is consistently assumed to be also of the power-law form but harder, with the 
index of  about (-2).  Such a simply spectrum is supposed to work in the ankle region (and below), but for 
higher energies the cut off also exists, what is clearly visible in both discussed experiment spectra and shown in Fig.\ref{paota}.
For the UHECR EG particles, if there are protons, nuclei or even some exotic particles like photons (which case we do not 
discussed in the present work), there exit well known physical mechanisms limiting their energies when they reach the Earth.  
In the case of protons the famous GZK cut-off \cite{PhysRevLett.16.748, 1966JETPL...4...78Z} works above few 
times 10$^{19}$ eV. It is the consequence of 
the existence of  the $\Delta$ resonance just in the energy region of collisions of the UHECR  nucleons with 
photons of cosmic 
microwave background (CMB). Below the GZK cut-off energy there is expected the effect of $e^+e^-$ pair production
by the UHECR 
protons in the CMB. This effect is not as big and definite as GZK, but it should be taken into account for more detailed, e.g., 
mass composition, analysis.

\begin{figure}
 \includegraphics[width=8.5cm]{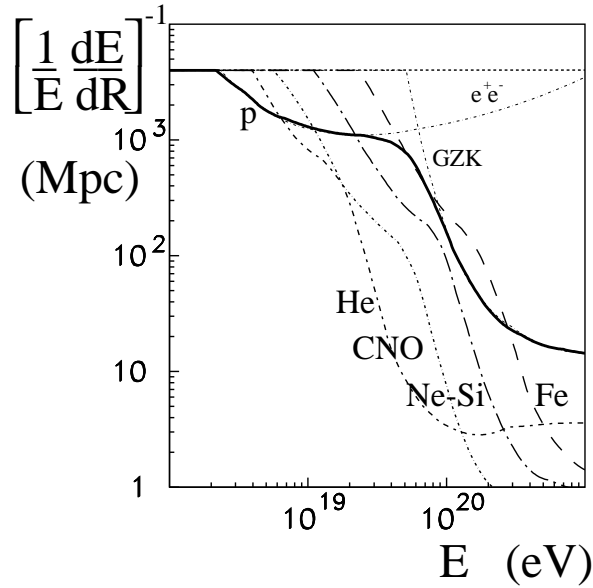}
\caption{The energy losses lengths for GZK and different nuclei photodisintegration processes. \label{gzk}}
\end{figure}

For nuclei the giant dipole resonance existence leads to fast disintegration of UHECR nuclei passing through
intergalactic photons of  CMB but also interacting with the intergalactic photon field in infrared (IR) range. 
Unfortunately the IR light intensity (and the IR photon energy spectrum) is not known as well as the CMB radiation. 
The detail shape of the nuclei cut-off is a source of the theoretical uncertainty  \cite{taylorthesis, Hooper2007199}
at the energies below the ones when the CMB smashes all nuclei fast and definitively. 
The physical processes listed above allows us to determine more exactly the shape of "the end of the cosmic ray spectrum":

The formula describing the EG UHECR flux adopted in the present analysis is shown in Eq.(\ref{feg}).
 
 \begin{equation}
                 \Phi_{\rm EG} (E)         {d \Phi_{\rm EG} \over dE}= {  {\Phi_0}_{\rm EG}} 
\:E^{-\gamma_{\rm EG}}\: \times \: \left\{ {
\begin{array}{ll}
f_{\rm GZK} & 
{\rm for\ protons}\\
                                                    f_{\rm nuc}  &
{\rm for\ nuclei  }
\end{array} 
}\right.
\label{feg}
\end{equation}
\noindent
where $f_{\rm GZK}(E)$  and $f_{\rm nuc}(E)$ are factors obtained numerically for the process of the energy 
losses of nucleons and nuclei which we shown in Fig.\ref{gzk}. The energy loss lengths are calculated
separately for protons and nuclei of the main four groups named as \he, \cno, \nesi, and \fe,
which contributions will be 
examined later on.

To get numerical values of the $f$ factors in Eq.(\ref{feg}) the CR flux reaching us from the sources have to be integrated  over their distribution in the space, which do not have to be uniform. We will come back to this point in Sec. \ref{distribution}.

\section{'World average' UHECR spectrum} 

All problems mentioned above make the situation at the "end of the spectrum" complicated, but we will try to answer all questions successfully.
We will start from combining both spectra (PAO and TA) together, removing the mentioned above experimental biases. 

Our analysis is based on the supposition that `the ankle' is the real feature of the observed UHECR flux, and the flux is
given by $\Phi_{\rm UHECR} = \Phi_{\rm G}+\Phi_{\rm EG}$.

From the experimental point of view UHECR flux has been examined by many group, historically, from the important John Linsley measurement in Volcano Ranch \cite{PhysRevLett_10_146}, but actually mostly by PAO and TA, each 
having its own energy resolution and possible biases in estimated energy and collecting aperture.
Both big contemporary working experiments are very complicated and the procedure of single shower energy determination is known in general, but there
are plenty of instrumental and analysis details which are not widely known. Some of them are described in more or less technical papers \cite{1475-7516-2017-04-038, ThePierreAugerCollaboration2015172}. The 'concluding' experimental spectra are 'checked' finally with respective 
Monte Carlo. This, however, adds the MC code, and simulation approximation uncertainties to the overall uncertainty of the final results. 
There is no way to overcome all this, and we have to live with it. 

The flux normalization discrepancy (vertical displacement) is also observed in the data (Fig.\ref{paota}), but it does not disturb our further composition analysis and conclusions.
The energy bias is also simply factor, which the energy should be multiplied by. It does not change the shape 
of the spectrum, too. But the experimental energy resolution  
changes the theoretical expectations for each experiment differently.
According to the
experiment group reports the energy resolution is similar for PO and TA and it is of order 0.1  in the $\log_{10}$ energy scale
\cite{paospectrum, taspesumm}.
 In particular, of course, it depends on the actual  sample of showers analyzed, the quality of events, and of the energy itself. 

According to the simulations, e.g.  \cite{Takeda2003447}, we assumed the energy reconstruction uncertainty
to be Gaussian on the logarithmic scale. Transferring the UHECR universal flux reaching the Earth to the spectrum measured by the   $j$-th 
experiment we have 
\begin{equation}
                {\Phi_{\rm obs}}_{j}(E) =% { d \Phi_{\rm obs}(E) \over d E}=
                 \int 
d (\log E')\: E'\: \Phi_{\rm UHECR}(E') \:
{1\over {\sqrt{2 \pi} \sigma_{j} }}\:
\exp \left[ - {\left({\log (E) - \log (E' )}\right)^2 \over \  2  {\sigma}_{j} ^2}  \right]~~~.
 \label{fold}
\end{equation}
For the exact power-law UHECR spectrum $\Phi_{\rm UHECR} (E)=E^{- \gamma}$ the effect of the folding it with the Gaussian 
resolution (on the $\log (E)$) does not change the shape, it remains power-law. It shifts only the spectrum
up  by the factor of $\exp(\gamma^2 \sigma_j^4 /2) $ on the flux intensity scale. However, if the spectrum has a more complicated structure it is eventually changed. The effect is obvious and was discussed already in \cite{0305-4616-14-6-020} and by us in 
\cite{Szabelski2002125}.  It makes the structures,
generally
smoothed. In the case of 'the ankle' it makes it less significant.
When we combined  spectra of many different experiments, 
each with its own, different resolution, this effect is very important and have to be taken into account if we wish to resolve 
the correct 'world average' UHECR
spectrum. For the present analysis, we are going to  combine only two spectra obtained by two EAS arrays with similar 
resolutions, the situation is simpler, but the effect still exists, what should be remembered.

The concept of the procedure is based, as it has been said, on the assumption, that the structure of 'the ankle' -- the dip between 
the $10^{18}$ eV and  $10^{19}$ eV, is the real phenomenon.
Below the energy of $10^{19}$ eV the cut-off factors of the EG component shown in Eq.(\ref{feg}) do not contribute.
If we, at the first step of combining PAO and TA spectra, limit the fitting procedure to the energies below $2 \times 10^{19}$ 
using not all points measured by PAO and TA,  
the fallowing  four parameters of the Galactic component in Eq.(\ref{fg}) can be determined:
\begin{itemize}
\item[-]
normalization $  {\Phi_0}_{\rm G}$
\item[-]
index $\gamma_{\rm G}$
\item[-]
position of the confinement cut-off $E_{\rm Gcut}$
\item[-]
width of the confinement cut-off $\sigma_{\rm G}$
\end{itemize}

\noindent
and two  Extragalactc component parameters describing the EG power-law in Eq.(\ref{feg}):
\begin{itemize}
\item[-]
normalization  $  {\Phi_0}_{\rm EG}$
\item[-]
index $\gamma_{\rm EG}$
\end{itemize}

\noindent
%At this stage we do not introduce the EG component cut-off. 
The parameters listed above, however, are not all parameters acting in the 'ankle' region. Two energy resolutions for both 
experiments together with one independent energy bias makes (if we took PAO energy measurement as a datum 
we set energy bias for PAO points to 0, and accordingly with the TA energy normalization) all together 9 parameters
 to be adjusted 
to the spectra of the 'ankle' measured by PAO (${Phi_{\rm PAO}}_i$) and TA (${\Phi_{\rm TA}}_i$). 
The spectra in the energy range of interest  (log(E) less than 19.2) consist of
16 points from PAO and TA data sets, each with its individual statistical uncertainty ($\Delta_i$).
The cost function which we try to minimize is the $\chi^2$ defined conventionally as

\begin{equation}
\chi ^2 = \sum_{\rm PAO}        (({\Phi_{\rm obs}}_{\rm PAO} (E_i) -  {\Phi_{\rm PAO}}_i)/{\Delta_{\rm PAO}}_i)^2
+ \sum_{\rm TA}     
   (({\Phi_{\rm obs}}_{\rm TA} (E_i) -  {\Phi_{\rm TA}}_i)/{\Delta_{\rm TA}}_i)^2 .
   \label{chi}
\end{equation}

\begin{figure}
\centerline{
 \includegraphics[width=8.5cm]{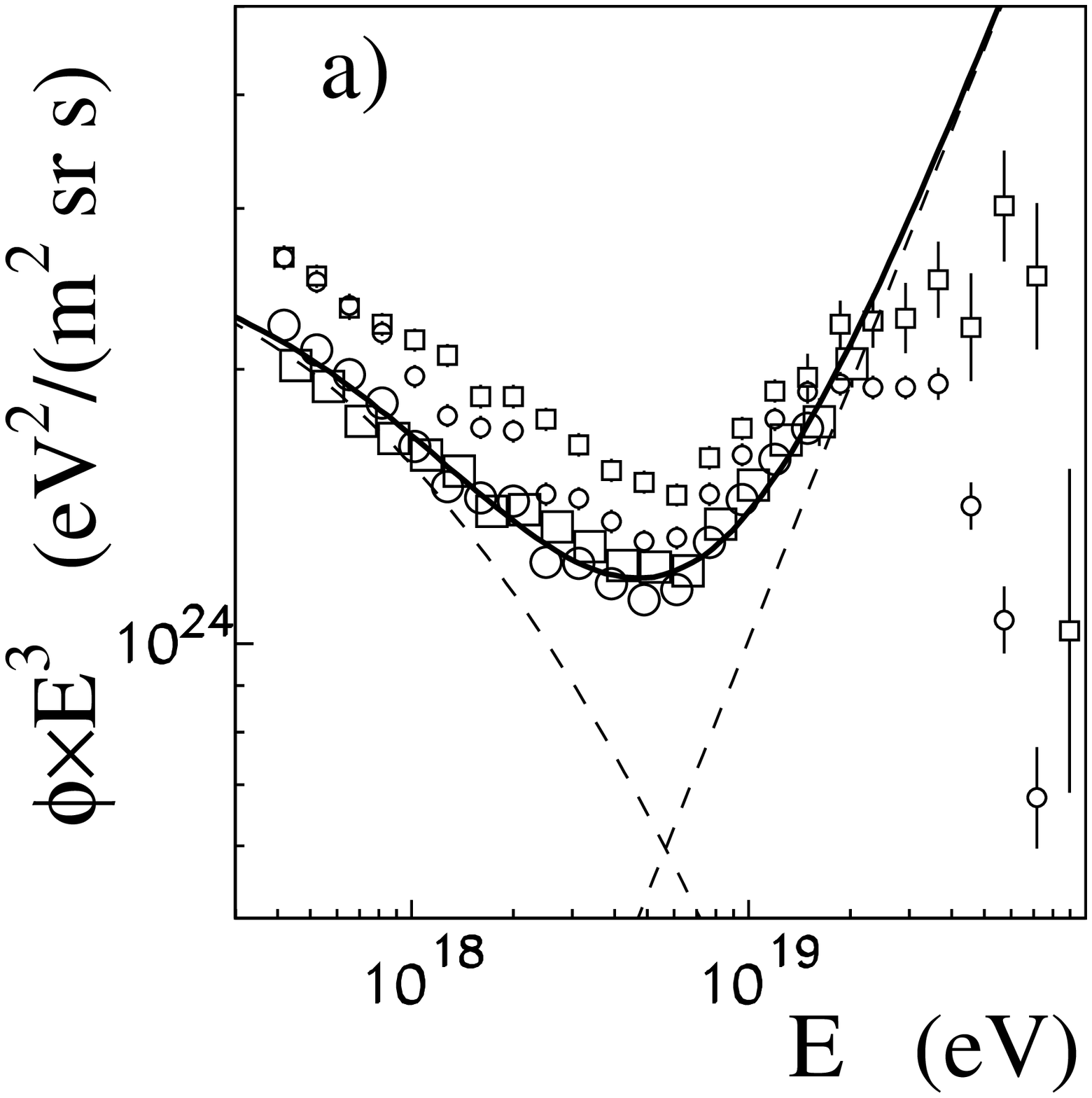}
\includegraphics[width=8.5cm]{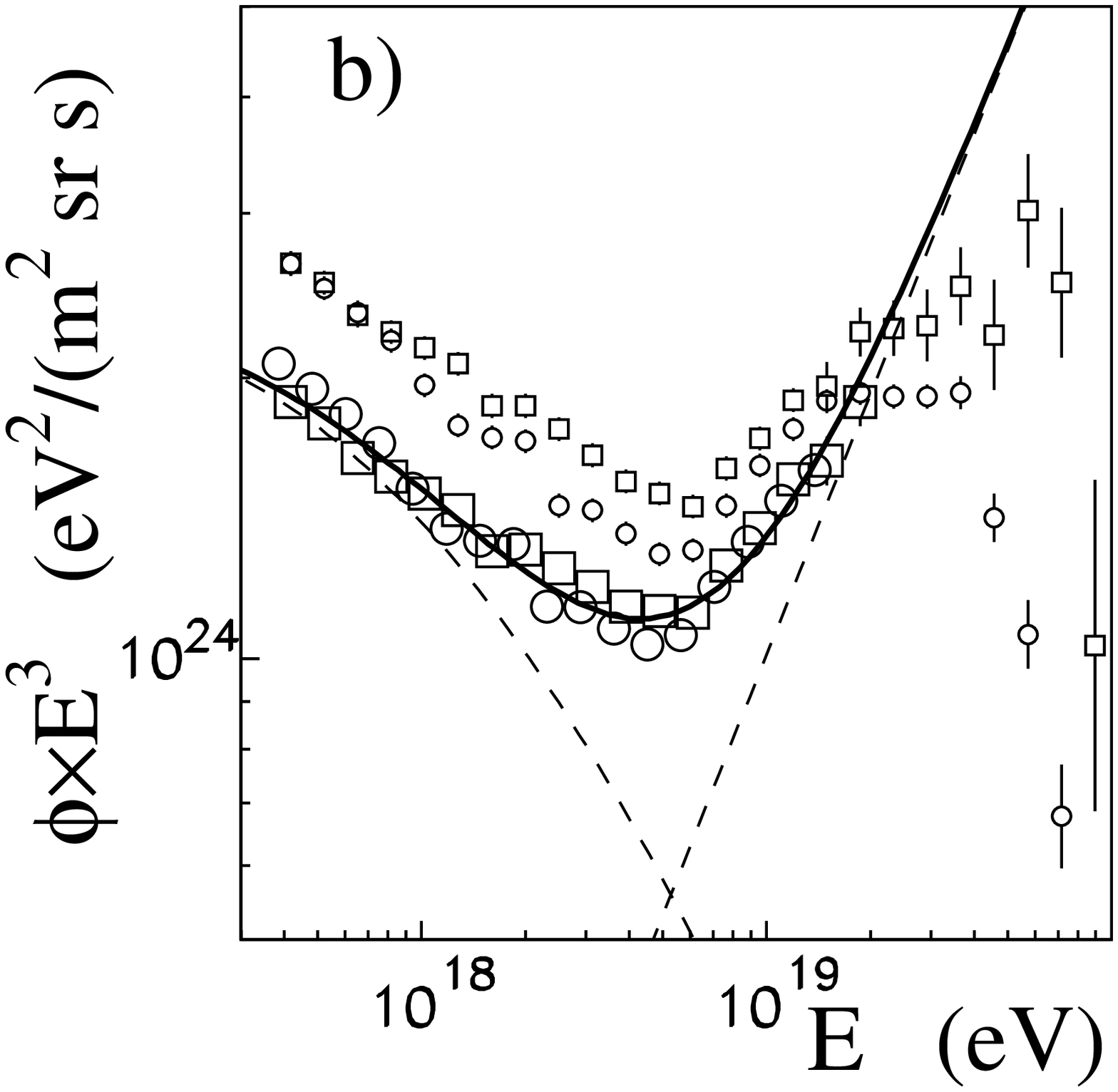}}
\caption{Energy spectrum fits of the G and pure power-law EG components 
in the 'ankle' energy region. 
On the left plot - a), the absolute normalization reported by PAO is used and the systematic shifts are applied only for TA data.
For the right plot - b), the opposite: the TA normalization is fixed and PAO is systematically shifted. 
The small symbols (circles for PAO and squares for TA)  are the original measured spectra. The big symbols are the 'unfolded' spectra using the logarithmic 
slope
of the model spectrum shown by solid black line. \label{fit_norm}}
\end{figure}

Of course not all parameters are independent and the procedure of $\chi^2$ minimization had to be made carefully consecutively in different parameter subspaces
omitting local minima and divergences. Eventually we obtained two fits to the 'ankle' spectra for the PAO and the TA energy normalizations.
They are showed in Fig.\ref{fit_norm}. Each plot shows the universal UHECR spectrum adjusted to both sets of data but for different experiment energy estimation used as a datum. The original measured data points are shown by small symbols (circles for PAO, squares for TA) and the solid line represents the unfolded spectrum. The big symbols are the  measured points 'corrected' by the experimental resolution of Eq.(\ref{fold}). The bias on the energy scale is set 0 for PAO points at the left plot, Fig.\ref{fit_norm}a, and 0 for TA on the right one, Fig.\ref{fit_norm}b. Dashed lines shows Galactic and Extragalactic components.

\begin{figure}
\centerline{
\includegraphics[width=8.5cm]{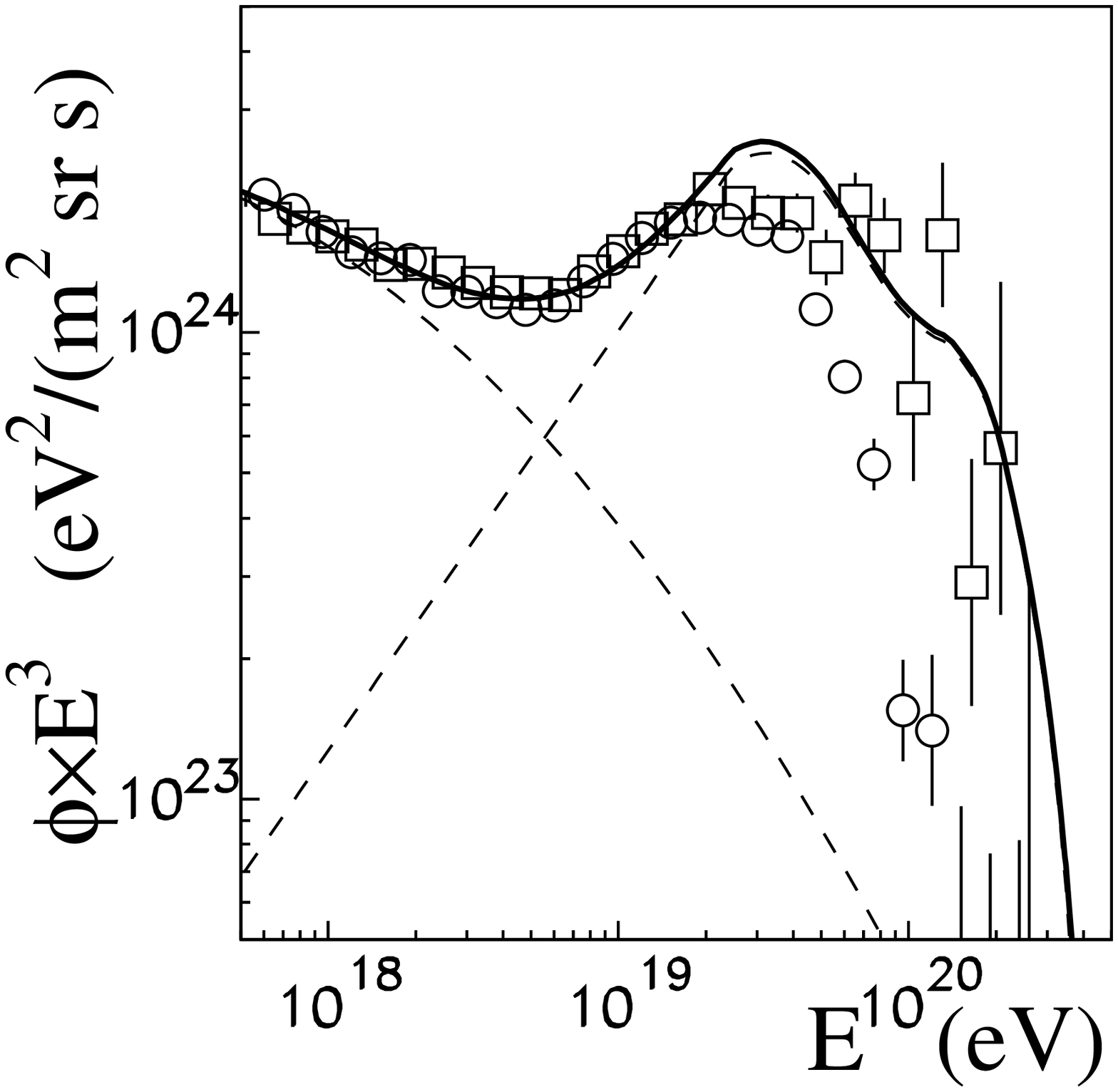}
\includegraphics[width=8.5cm]{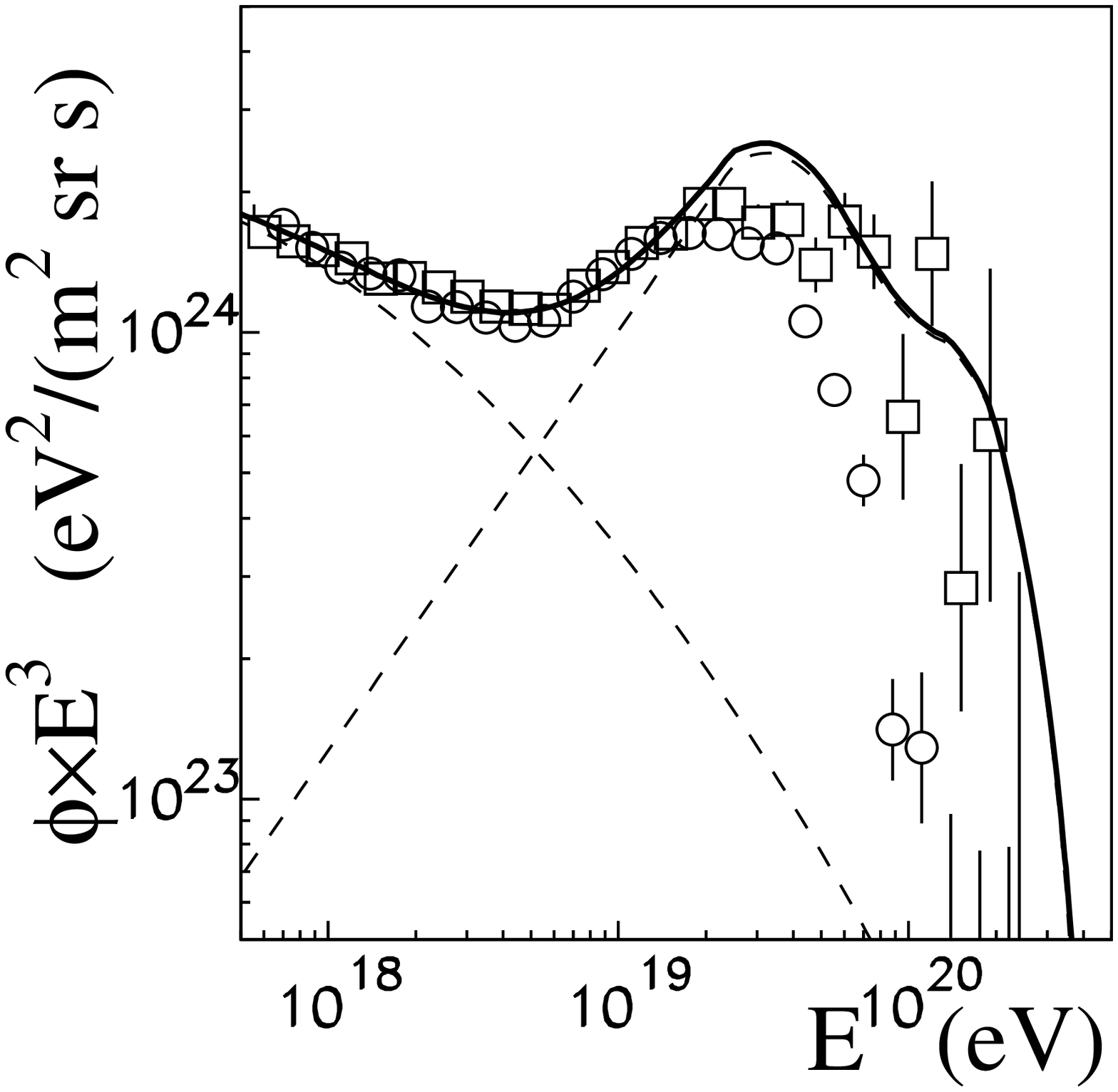}}
\caption{Fits of our model parameters in the whole energy range to the spectra measured by PAO and TA. 
Left and right plots are as in Fig.\ref{fit_norm}. 
The symbols are 'unfolded' with the universal spectrum shown as a black solid lines. The EG component is of the form of power-law with the high energy 
cut described in the text.
 \label{fit_cut}}
\end{figure}

With the values of the model parameters found for the uncut EG component, we have calculated the universal UHECR spectrum 
$\Phi_{\rm UHECR} = \Phi_{\rm G}+\Phi_{\rm EG}$ in the whole energy range. We used here, as an example for 'the end of the spectrum' cut-off the one 
calculated for the \fe \ group in the uniform CR source distribution. For other choices the results do not change significantly. This, more realistic spectrum shape  is needed to shift each measured point accordingly.
We did the calculations for both (PAO and TA) energy normalizations to get the local value of the 
logarithmic slope of the spectrum ($\gamma$) needed to unfold the measured spectra.
 It does not depend much on the particular choice of the cut-off shape (within  reasonable limits) and the Fig.\ref{fit_norm} presents the 
situation for pure iron nuclei flux and uniform distribution of CR sources. More detail will 
be shown and discussed below in Sec. \ref{composition}.
The results of the unfolding procedure is shown in Fig.\ref{fit_cut}. The 'universal UHECR spectra' are shown by the solid lines for both normalization cases: Fig.\ref{fit_cut}a for PAO and Fig.\ref{fit_cut}b for TA, respectively. Points (circles for PAO and squares for TA) represent 'unfolded data points' of both experiments.

\begin{figure}
\centerline{
 \includegraphics[width=8.5cm]{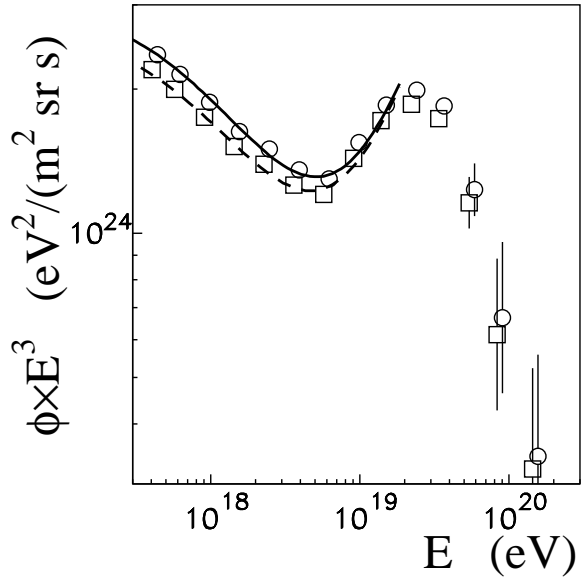}
}
\caption{The 'world average' UHECR spectrum. The circles and squares are combined PAO and TA results with the PAO and TA normalization used, respectively. The solid line is the ankle part of the universal UHECR spectrum with PAO normalization, the dashed line shows the TA normalization fit to the ankle. \label{ave}}
\end{figure}

As we can see in Figs.\ref{fit_norm} and \ref{fit_cut} the 'unfolded' points from both measurements lie closely to the 
proposed
universal UHECR spectrum, at least in the 'ankle' region. There is seen the small difference between the PAO and TA energy  normalizations (left and right plots in Figs.\ref{fit_norm} and \ref{fit_cut}), 
but here the difference is only a systematic shift and we can combine the point from both experiments together. It has been done and the resulting 
universal 'world average'  UHECR spectrum is shown in Fig.\ref{ave} and in Table \ref{univ}.

%\begin{table}
%\caption{\label{univ}Universal UHECR spectrum for two energy normalizations.}
%%\begin{ruledtabular}
%\begin{tabular}{rcl|crcl|crcl}
 %\multicolumn{3}{c|}{ }&\multicolumn{4}{c|}{PAO normalization}&\multicolumn{4}{c}{TA normalization}\\
%\hline
% \multicolumn{3}{c|}{Energy bin}&Energy&\multicolumn{3}{c|}{Flux $\times E^3$}&Energy&\multicolumn{3}{c}{Flux $\times E^3$ }\\
%\hline
%17.4&-&17.6 &17.57&24.307&$\pm$&0.016&17.55&24.282&$\pm$&0.001\\
%17.6 &-&17.8&17.72&24.266&$\pm$&0.014&17.70&24.241&$\pm$&0.008\\
%17.8&-&18.0&17.92&24.208&$\pm$&0.011&17.90&24.183&$\pm$&0.011\\
%18.0&-&18.2&18.12&24.146&$\pm$&0.008&18.10&24.121&$\pm$&0.014\\
%18.2&-&18.4&18.32&24.108&$\pm$&0.013&18.30&24.083&$\pm$&0.019\\
%18.4&-&18.6&18.52&24.062&$\pm$&0.007&18.50&24.036&$\pm$&0.016\\
%18.6&-&18.8&18.72&24.041&$\pm$&0.006&18.70&24.013&$\pm$&0.016\\
%18.8&-&19.0&18.92&24.115&$\pm$&0.012&18.90&24.085&$\pm$&0.017\\
%19.0&-&19.2&19.10&24.193&$\pm$&0.007&19.09&24.157&$\pm$&0.013\\
%19.2&-&19.4&19.30&24.222&$\pm$&0.009&19.30&24.181&$\pm$&0.018\\
%19.4&-&19.6&19.49&24.173&$\pm$&0.005&19.49&24.136&$\pm$&0.013\\
%19.6&-&19.8&19.69&24.088&$\pm$&0.052&19.69&23.979&$\pm$&0.060\\
%19.8&-&20.0&19.88&23.756&$\pm$&0.157&19.87&23.730&$\pm$&0.166\\
%&$>$&20.0     &20.11&23.456&$\pm$&0.212&20.10&23.447&$\pm$&0.218
%\end{tabular}
%%\end{ruledtabular}
%\end{table}

\begin{table}
\caption{\label{univ}Universal UHECR spectrum for two energy normalizations.}
%\begin{ruledtabular}
\begin{tabular}{rcl|rcl|rcl}
 \multicolumn{3}{c|}{ }&\multicolumn{3}{c|}{\ \ \ PAO normalization\ \ \ }&\multicolumn{3}{c}{\ \ \ TA normalization\ \ \ }\\
\hline
 \multicolumn{3}{c|}{Energy bin}&\multicolumn{3}{c|}{log$_{10}$(Flux $\times E^3$)}&\multicolumn{3}{c}{\ \  log$_{10}$(Flux $\times E^3$) }\\
\hline
\ \ \ \ 17.4&-&17.6\ \ \ \ \ &\ \ \ \ \ 24.296&$\pm$&0.015 &\ \ \ \ \ \  24.264&$\pm$&0.015\\
17.6 &-&17.8&24.254&$\pm$&0.008&24.223&$\pm$&0.013\\
17.8&-&18.0&24.196&$\pm$&0.011&24.165&$\pm$&0.011\\
18.0&-&18.2&24.135&$\pm$&0.014&24.104&$\pm$&0.008\\
18.2&-&18.4&24.098&$\pm$&0.019&24.067&$\pm$&0.013\\
18.4&-&18.6&24.055&$\pm$&0.016&24.024&$\pm$&0.008\\
18.6&-&18.8&24.036&$\pm$&0.016&24.004&$\pm$&0.008\\
18.8&-&19.0&24.112&$\pm$&0.017&24.080&$\pm$&0.012\\
19.0&-&19.2&24.191&$\pm$&0.013&24.158&$\pm$&0.007\\
19.2&-&19.4&24.222&$\pm$&0.018&24.192&$\pm$&0.012\\
19.4&-&19.6&24.189&$\pm$&0.013&24.162&$\pm$&0.009\\
19.6&-&19.8&24.014&$\pm$&0.060&23.987&$\pm$&0.054\\
19.8&-&20.0&23.747&$\pm$&0.166&23.712&$\pm$&0.158\\
&$>$&20.0 &23.458&$\pm$&0.218&23.431&$\pm$&0.210
\end{tabular}
%\end{ruledtabular}
\end{table}

%=============================================================
\section{UHECR mass composition \label{composition}}
With the 'world average' UHECR spectrum we tried to perform the mass composition estimation of "the end of the cosmic ray spectrum".

With the number of the model parameters increased by the fractions of each kind of nuclei (we limited ourselves to five groups of nuclei only) for both: G and EG components 
we are dealing with almost 20 parameters space for the minimization. Is is obviously easy to reproduce the 14 points of the 'world average spectrum' even if 
some of the parameters are highly correlated. However, we have another two sets of observables of extensive air showers which are closely 
related to the mass composition: the $\langle X_{\rm max} \rangle$ - average depth of the shower development maximum and rms$_{X\rm max}$ - the width of the 
$X_{\rm max}$
distribution. PAO and TA published respective data \cite{PhysRevD.90.122005, Abbasi201549} and again some inconsistencies between PAO and TA data sets are seen. They do not allow us to make the 'world 
average'
by averaging points of both experiments. We are going to use the data as they are. The uncertainties concerning the 
Monte Carlo predictions \cite{PhysRevD.90.122006, PhysRevLett.104.161101, 1475-7516-2013-02-026}, differences between models of hadronic interactions, and possible experimental systematics make the analysis 
harder and sometimes 
even problematic, see e.g., \cite{PhysRevD.92.021302, PhysRevD.92.063011}. But in spite of all this there is the tendency in the observations and the general trends 
should be reproduced by the model to be accepted.
%------------------------------------------------------------------------------------------------------------------------------

\subsection{Distribution of the Extragalactic UHECR sources \label{distribution}} 

To perform the mass composition analysis we have to take into account as much as we can the possible effects of 
 processes leading to loss of cosmic ray particles energy throughout they path from the sources to us. 
As it was said the form of 'the end of cosmic ray spectrum' depends on the optical depth obtained by the integration of the energy losses with the distances CR travel which is weighted 
by the CR source distribution $d(R)$
\begin{equation}
               \tau = \int d R\: \left( {1 \over E}\:{dE \over dR} \right)\:  \Phi_{\rm source}(E) \:  d(R) ,
 \label{fnuc}
\end{equation}
where ${1 \over E} {dE \over dR}$ are shown in Fig.\ref{gzk}.

The most obvious, as a first approximation to be used, is the uniform CR source distribution. It is used in the recent PAO 
analysis \cite{1475-7516-2017-04-038}.
We will tested this possibility as the first try. 

\begin{figure}
\centerline{
 \includegraphics[width=6cm]{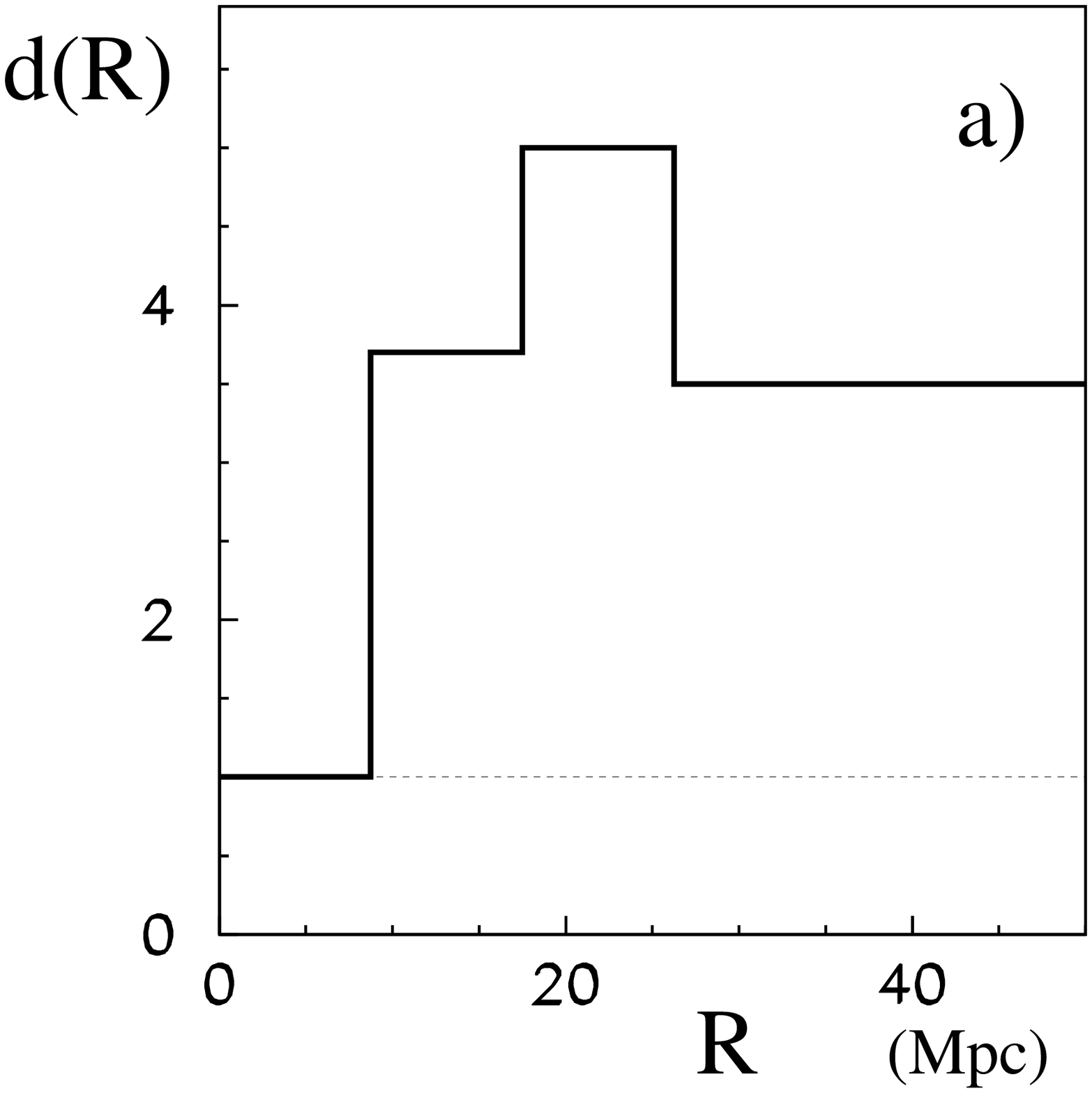}
 \includegraphics[width=6cm]{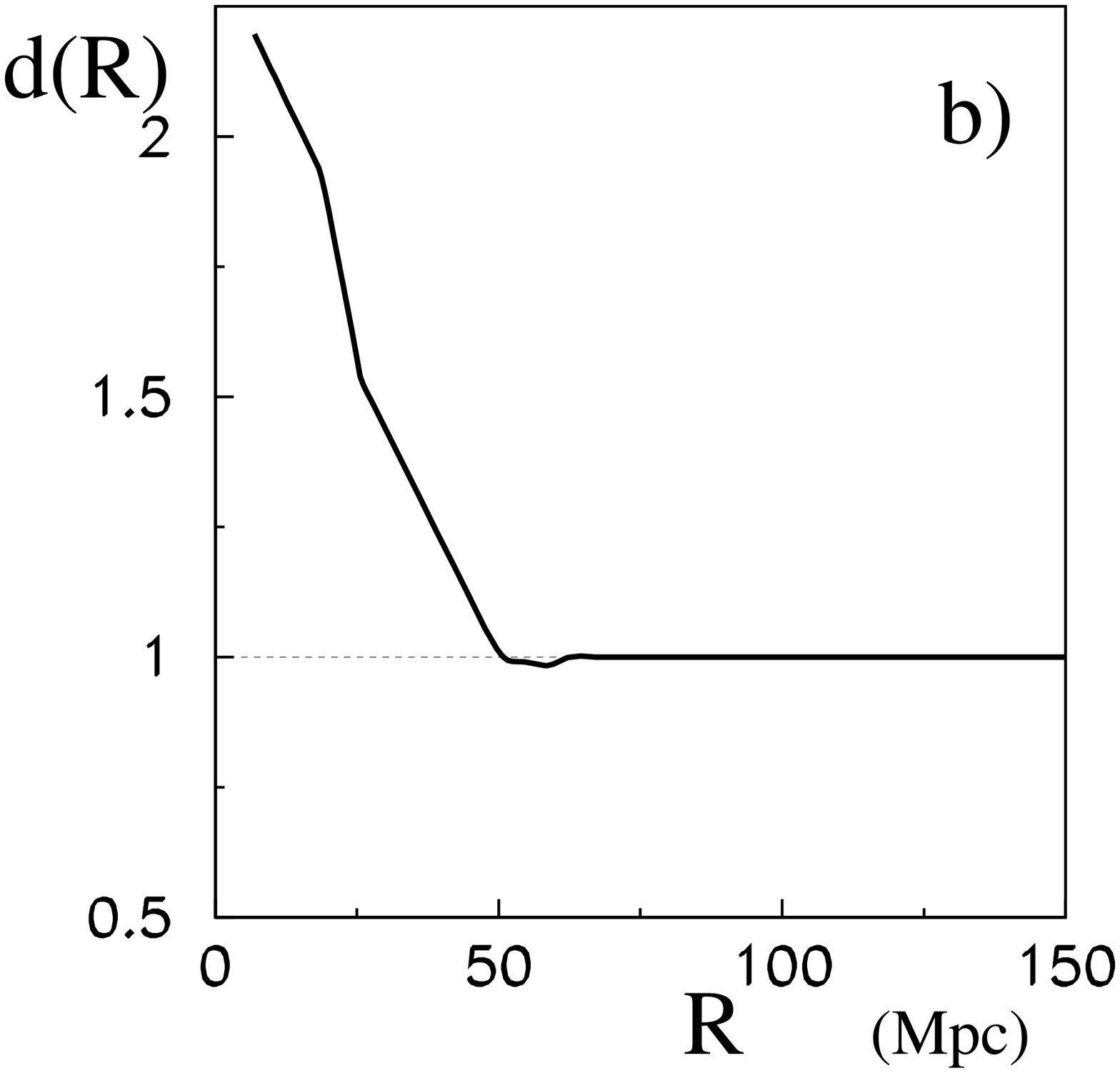}}
\centerline{
 \includegraphics[width=6cm]{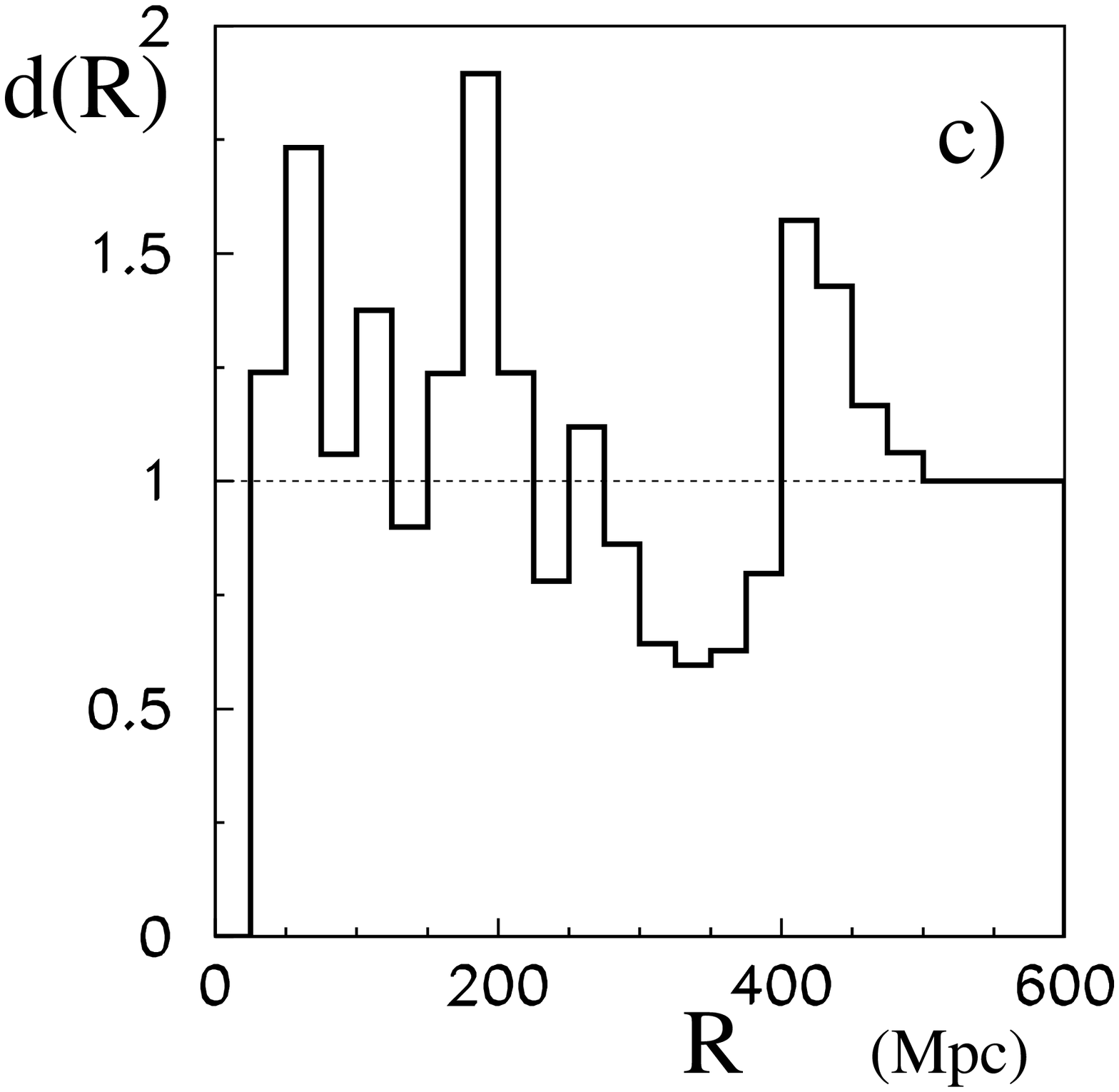}
 \includegraphics[width=6cm]{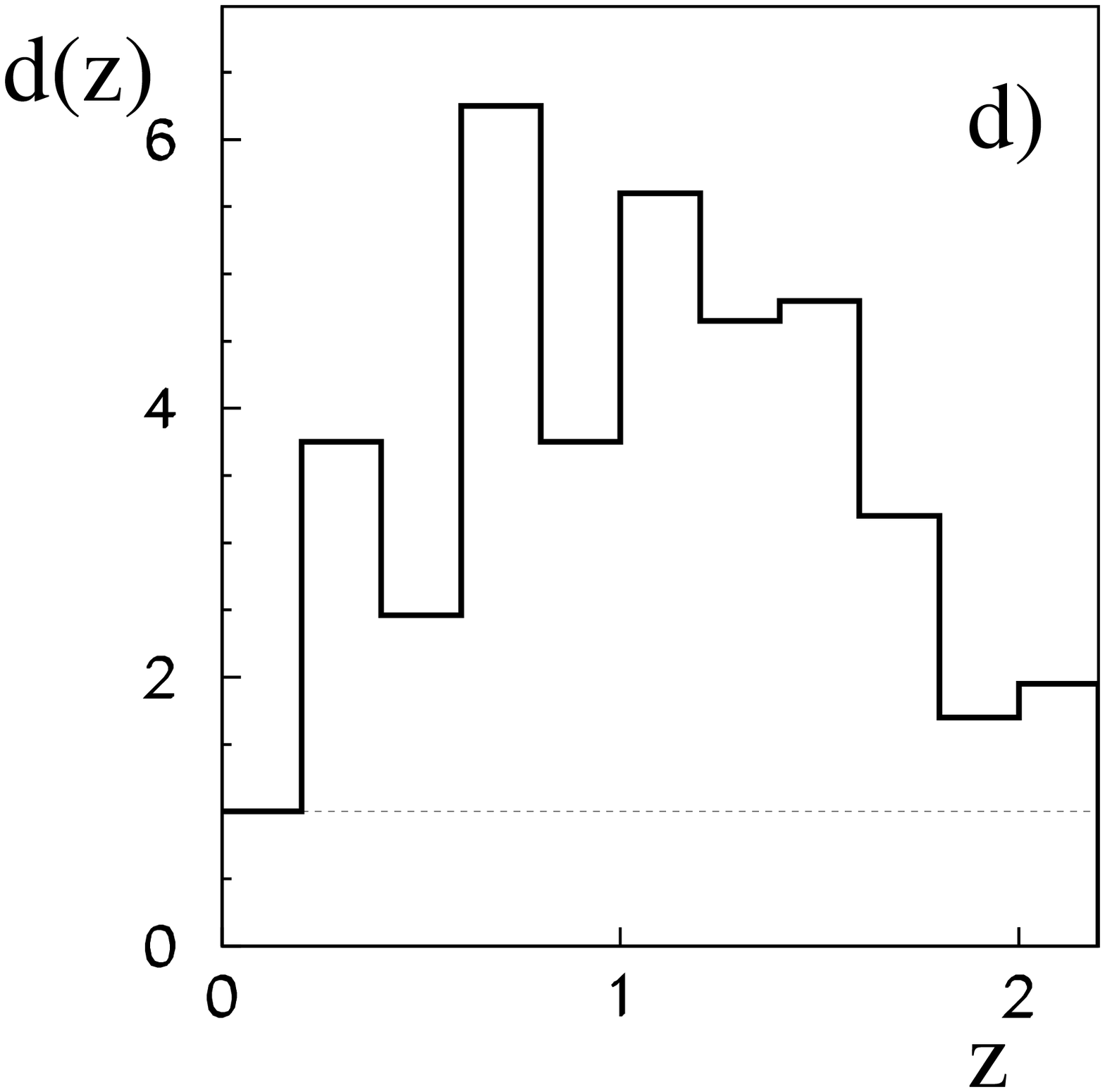}}
\centerline{
 \includegraphics[width=6cm]{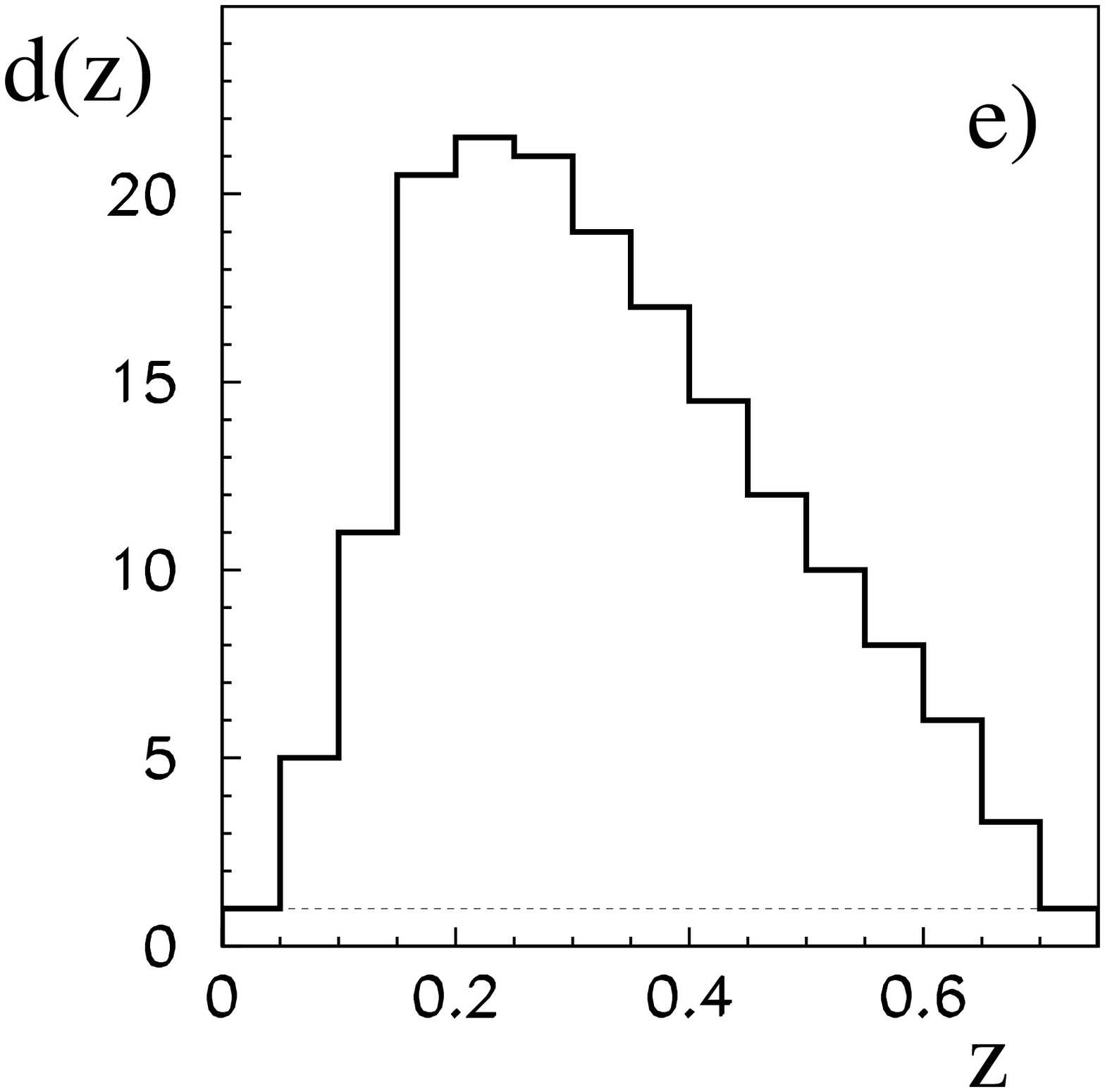}
 \includegraphics[width=6cm]{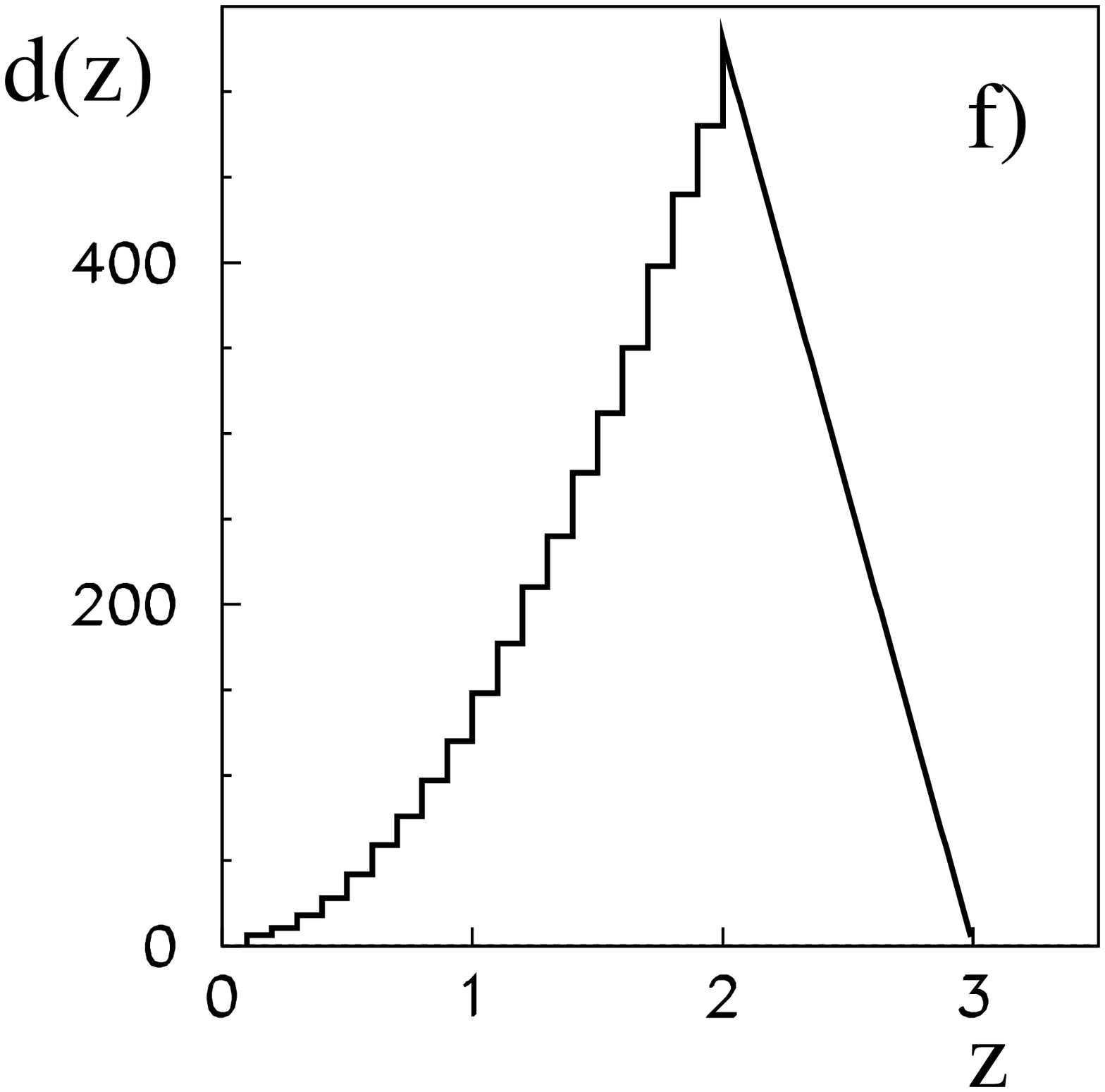}}

\caption{Distribution of UHECR sources analyzed in the present work: a) - colliding galaxies \cite{0954-3899-22-12-013} , b)  - Medina-Tanco proposition \cite{1538-4357-510-2-L91}, c) - DW \cite{dudarewicz}, 
e) - CDM \cite{wwmexico},  f) - galaxy clusters \cite{Bahcall:1988ch} and g) - quasar distributions. \label{dist}
}
\end{figure}

Looking more deeply, there are some constrains, mostly from the anisotropy measurement \cite{1475-7516-2012-04-040} 
that the number of UHECR sources close (within few Mpc from us) should be small, possibly even zero \cite{PhysRevD.74.043005}.  
So, our second guess is the uniform distribution but only for distances greater than 50 Mpc with no sources within. 
Such solution agree well with the observed lack of anisotropy.
The less drastic modification is the distribution of 'colliding galaxies' from \cite{0954-3899-22-12-013} presented in Fig.\ref{dist}a. 
Quite different is the proposition of Medina-Tanco in his analysis of the AGASA data \cite{1538-4357-510-2-L91} (shown in  Fig.\ref{dist}b) it is based of counting galaxies in the catalogue of CfA Redshift Survey from 1998.

Counting of galaxy clusters in the vicinity of our Galaxy gives another more realistic distribution of possible UHECR sources. We followed here the work of Dudarewicz and Wolfendale \cite{dudarewicz} which used maps from \cite{1992ApJ...388....9T}
and this distribution is shown in Fig.\ref{dist}c.

Modification on  larger scales in our analysis are %the gal......?????--- I can not identify the source of this distribution-----(Fig.\ref{dist}d) and 
represented by `the cold dark matter distribution' from Ref. \cite{wwmexico}
 (Fig.\ref{dist}d). Going further away we test the 
cosmological evolution models with galaxy clusters \cite{Bahcall:1988ch} (Fig.\ref{dist}e) and the strong cosmological evolution models with n$=5-6$ distribution motivated by distribution of quasars (Fig.\ref{dist}f).
All the distributions (with some others) were examined  by us in our earlier work \cite{0954-3899-34-9-003}. %all are shown in Fig\ref{dist}.

\begin{figure}
\centerline{
 \includegraphics[width=5.4cm]{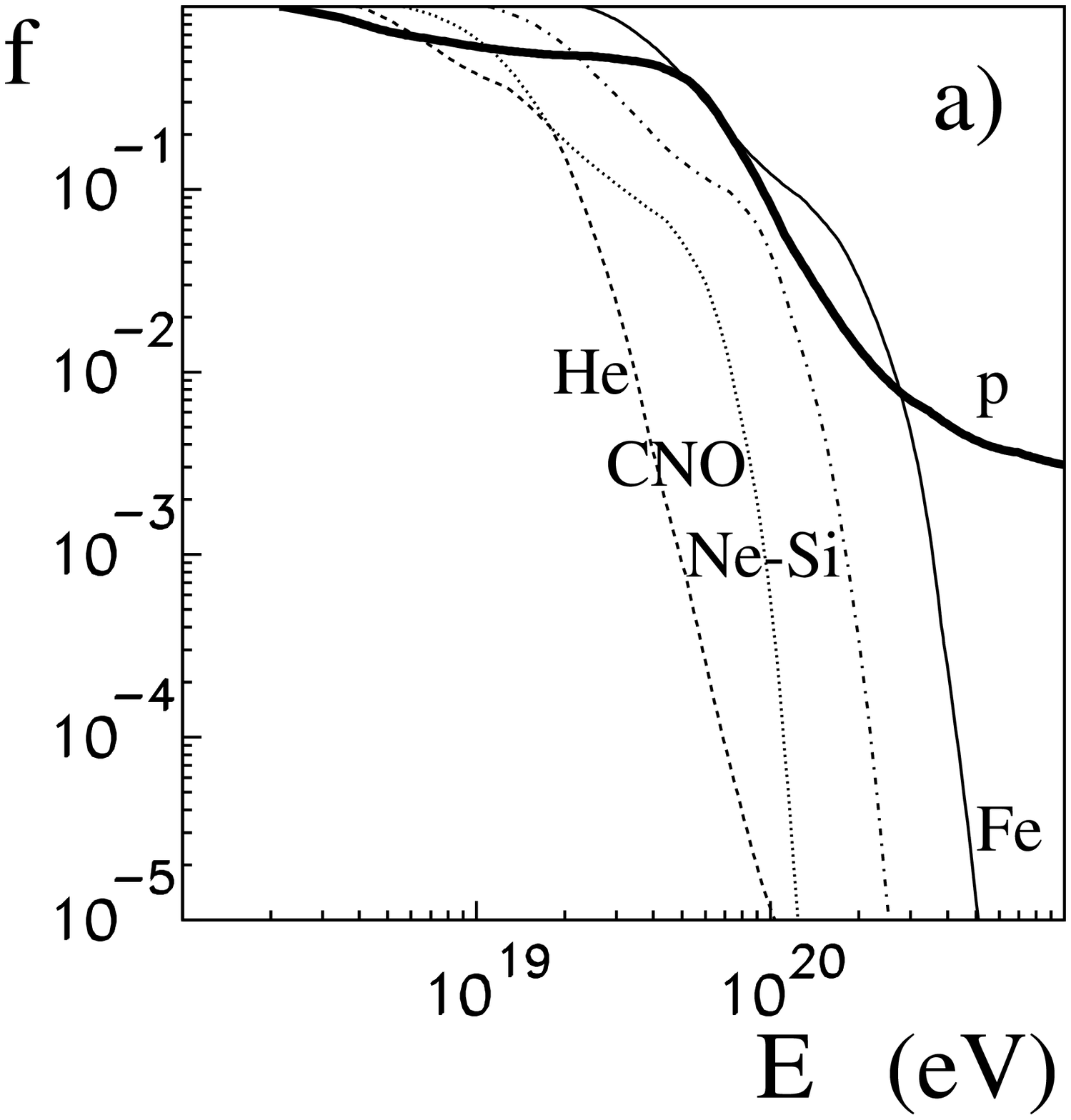}
 \includegraphics[width=5.4cm]{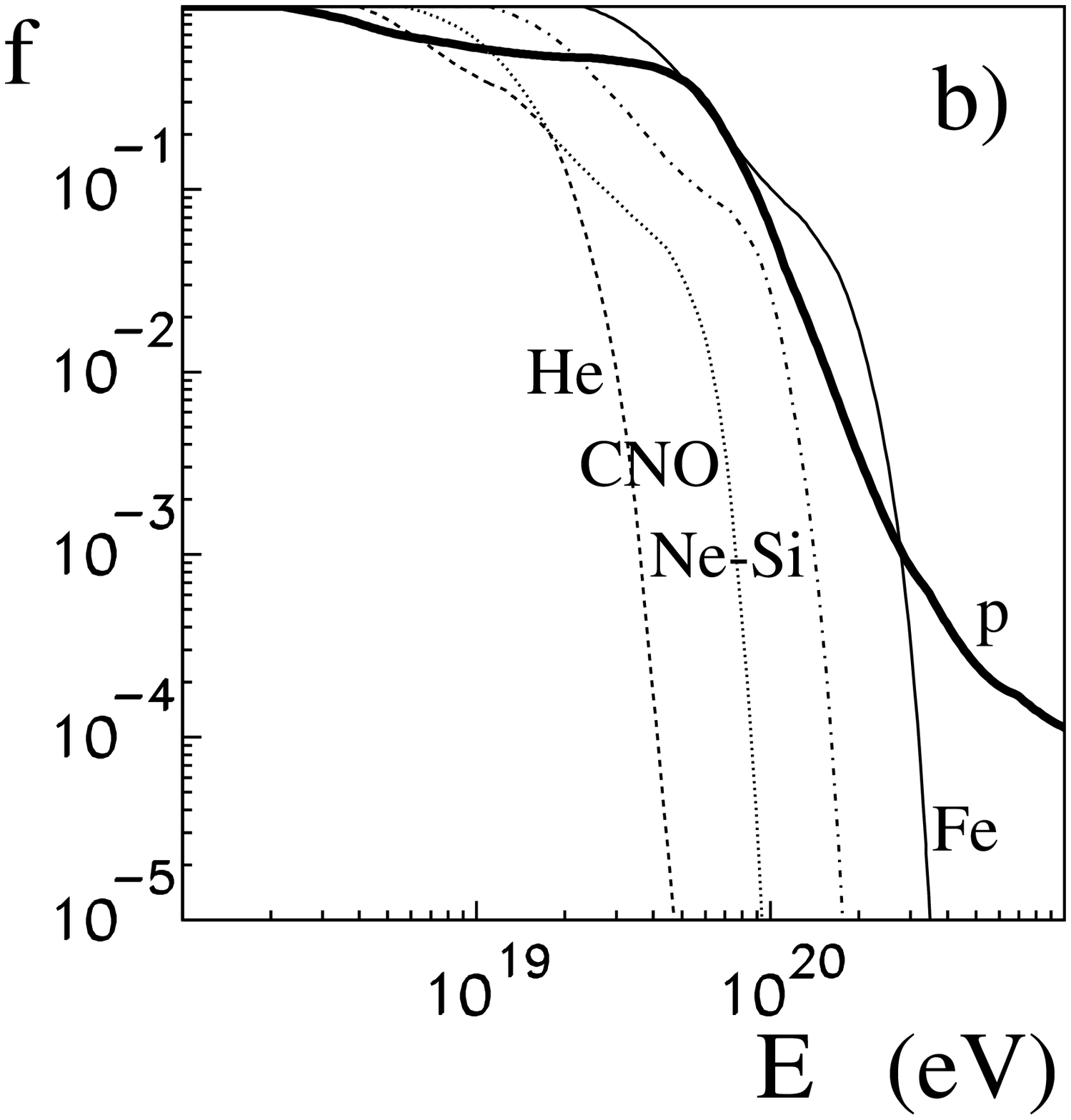}}
\centerline{
 \includegraphics[width=5.4cm]{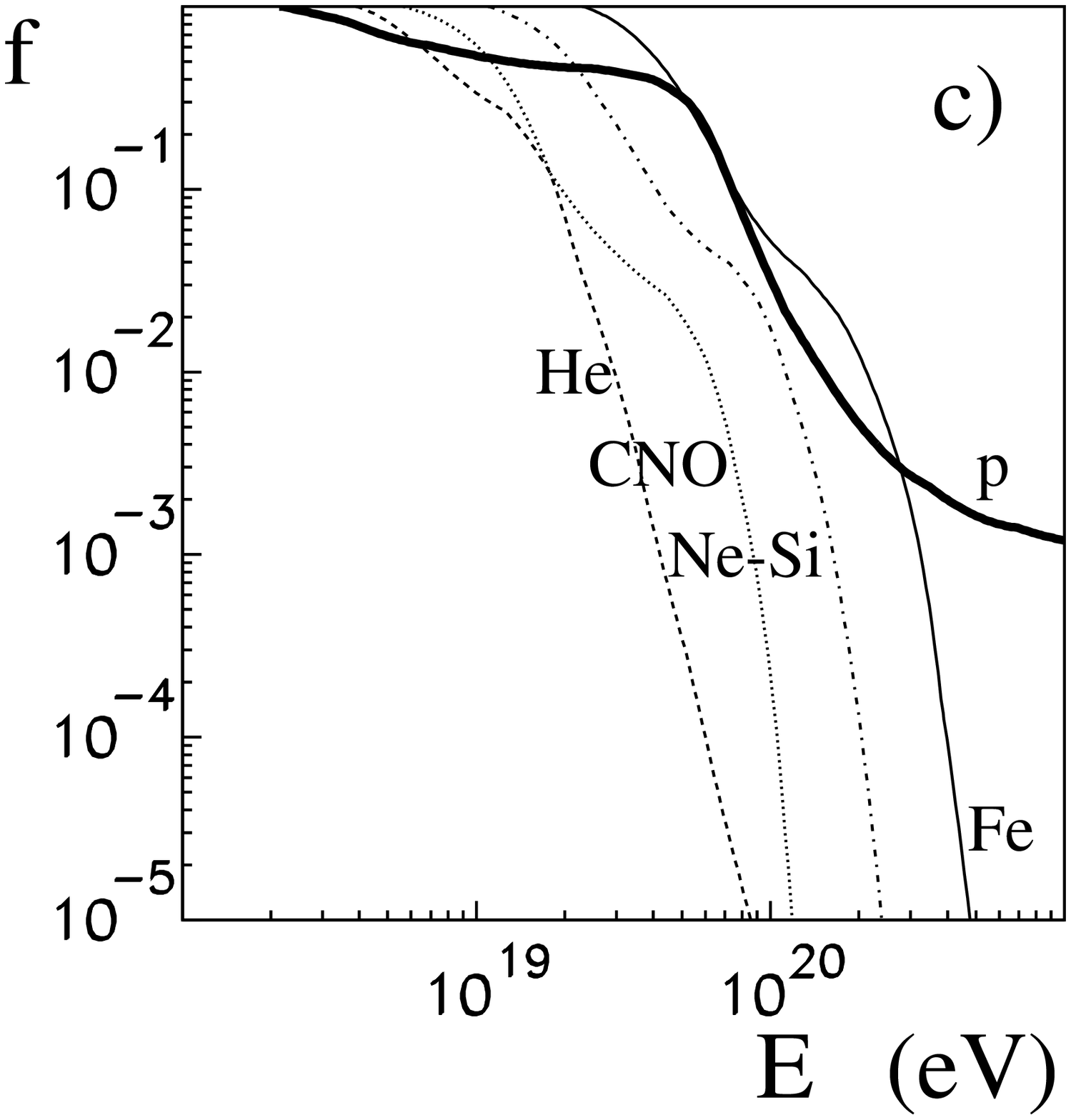}
 \includegraphics[width=5.4cm]{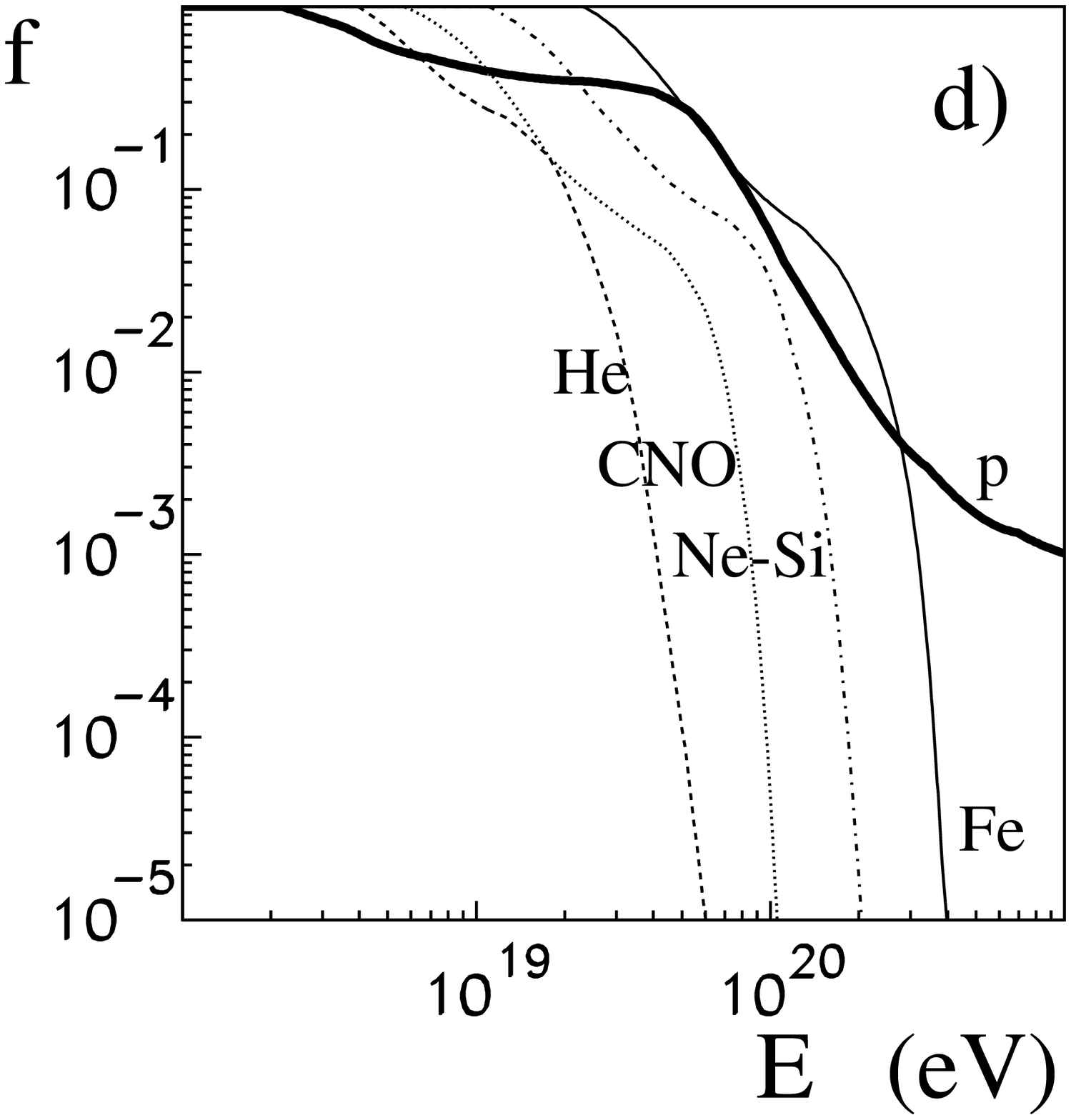}}
\centerline{
 \includegraphics[width=5.4cm]{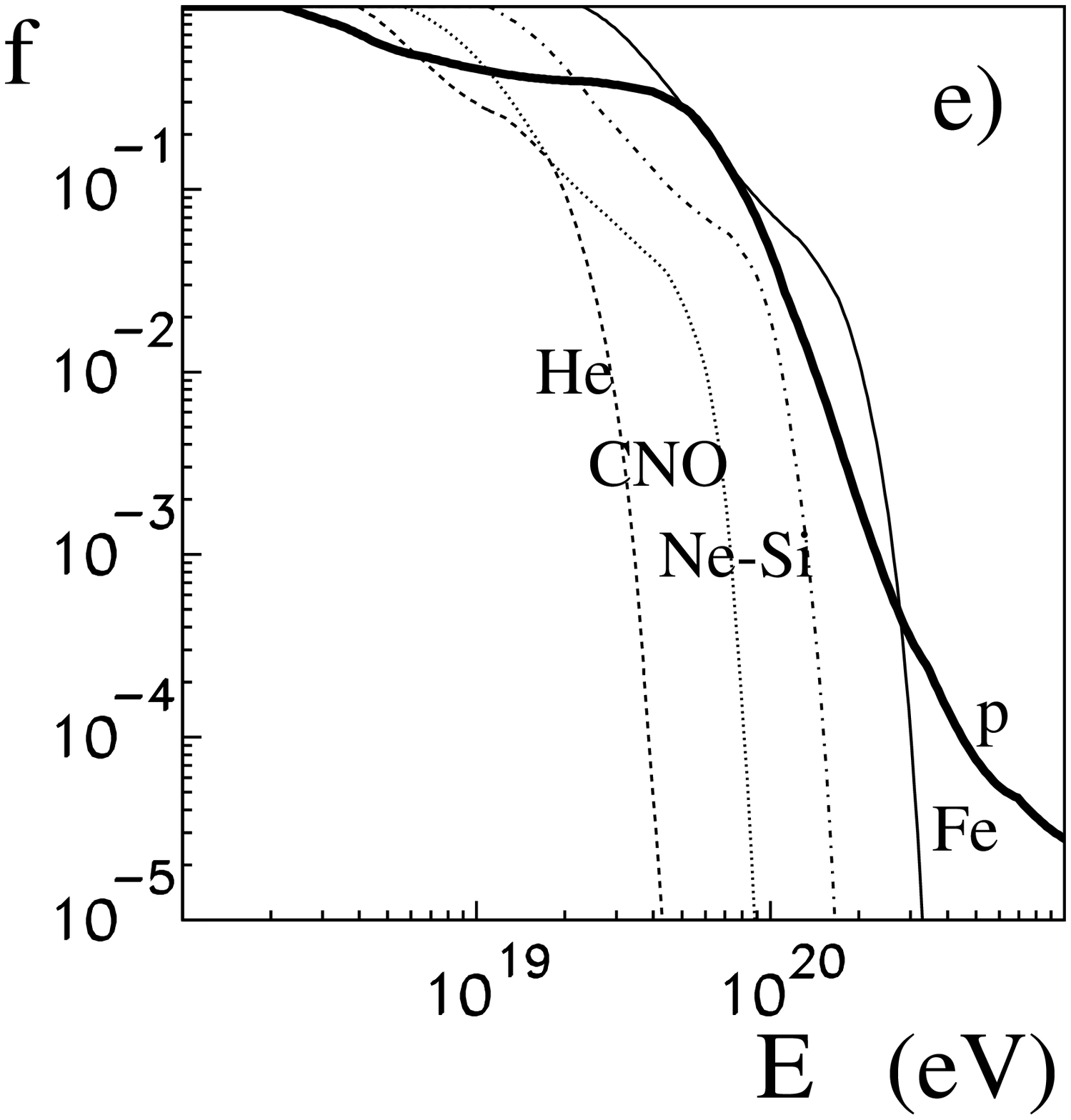}
 \includegraphics[width=5.4cm]{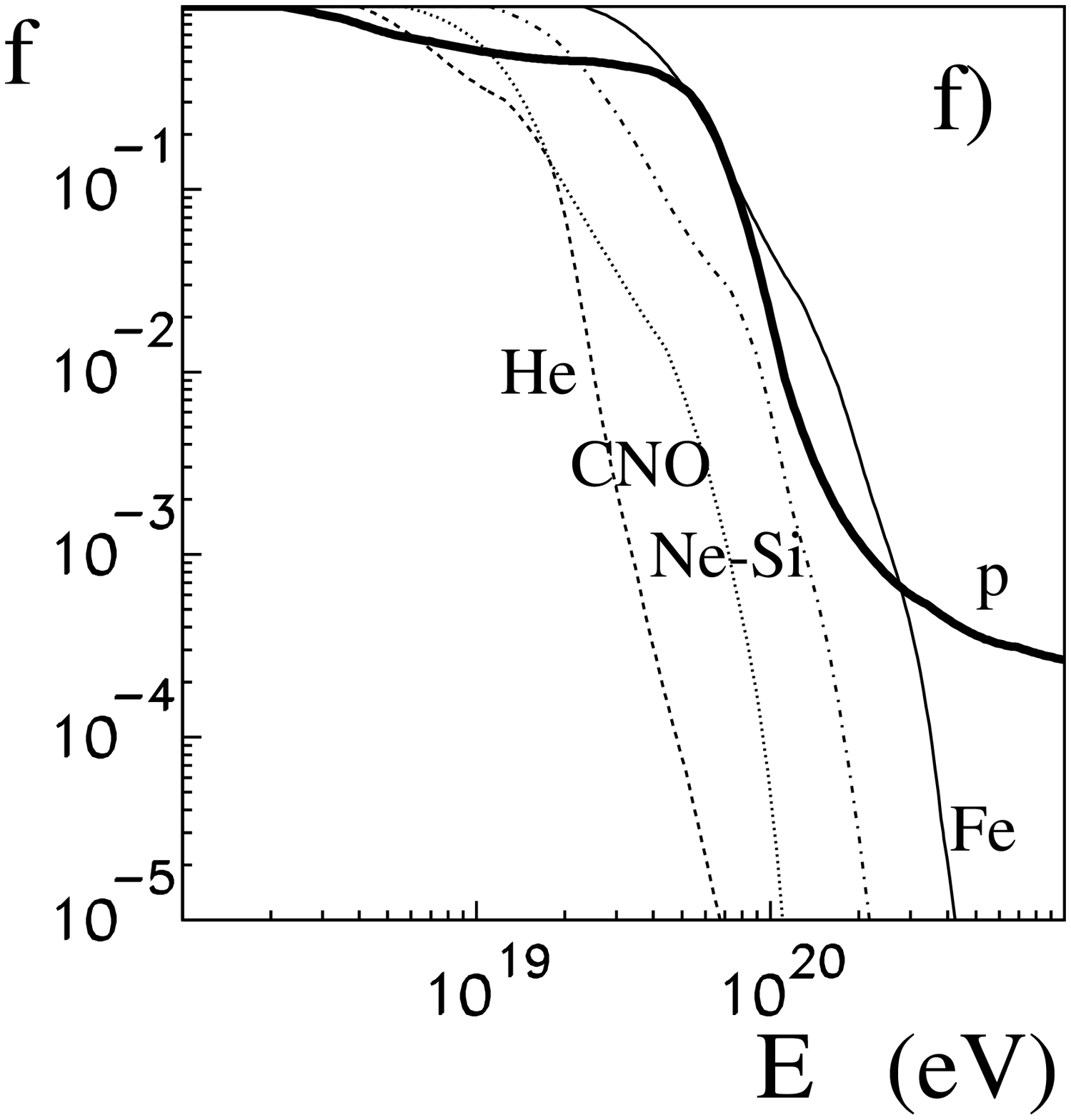}}
\centerline{ 
\includegraphics[width=5.4cm]{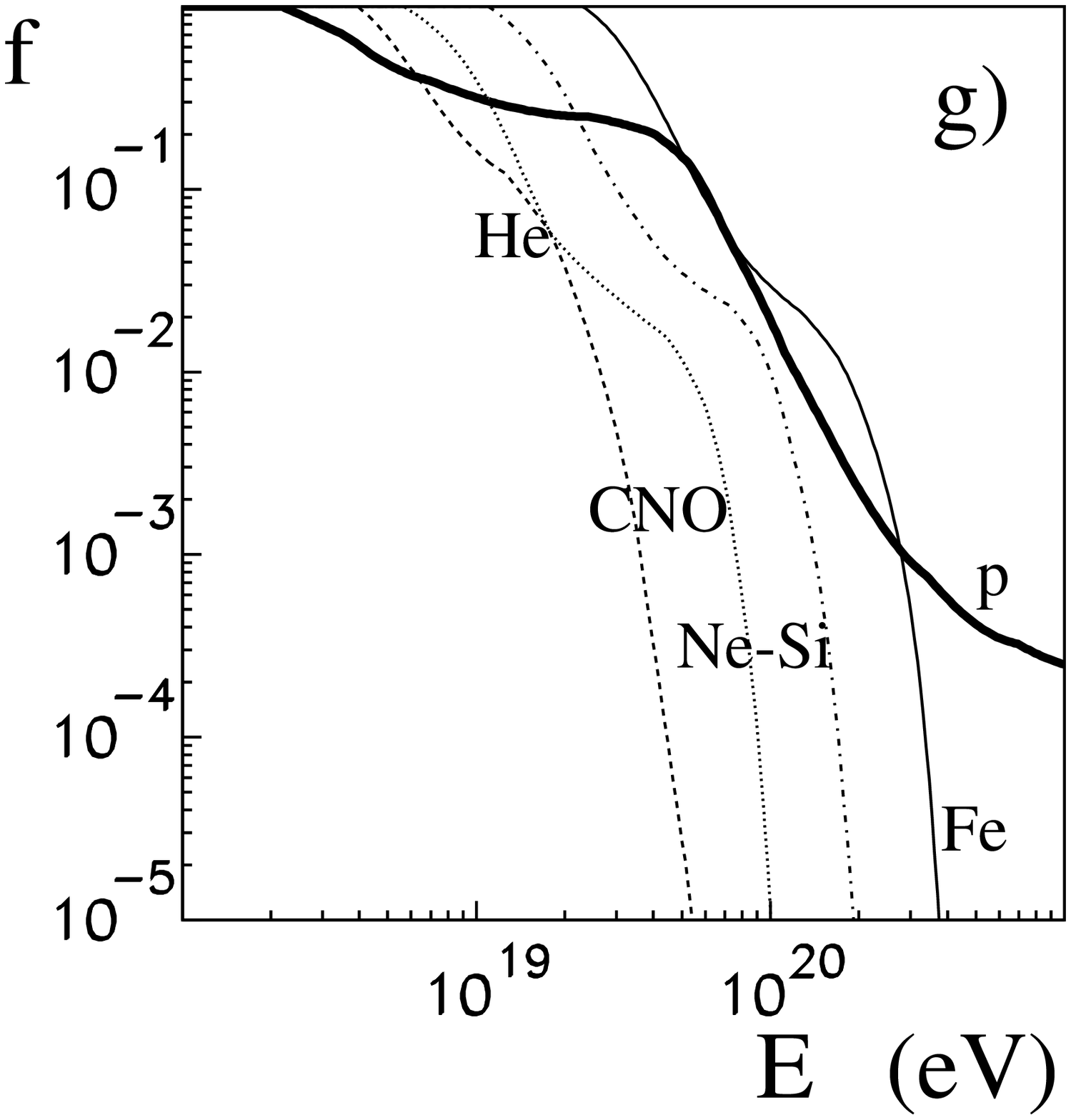}
 \includegraphics[width=5.4cm]{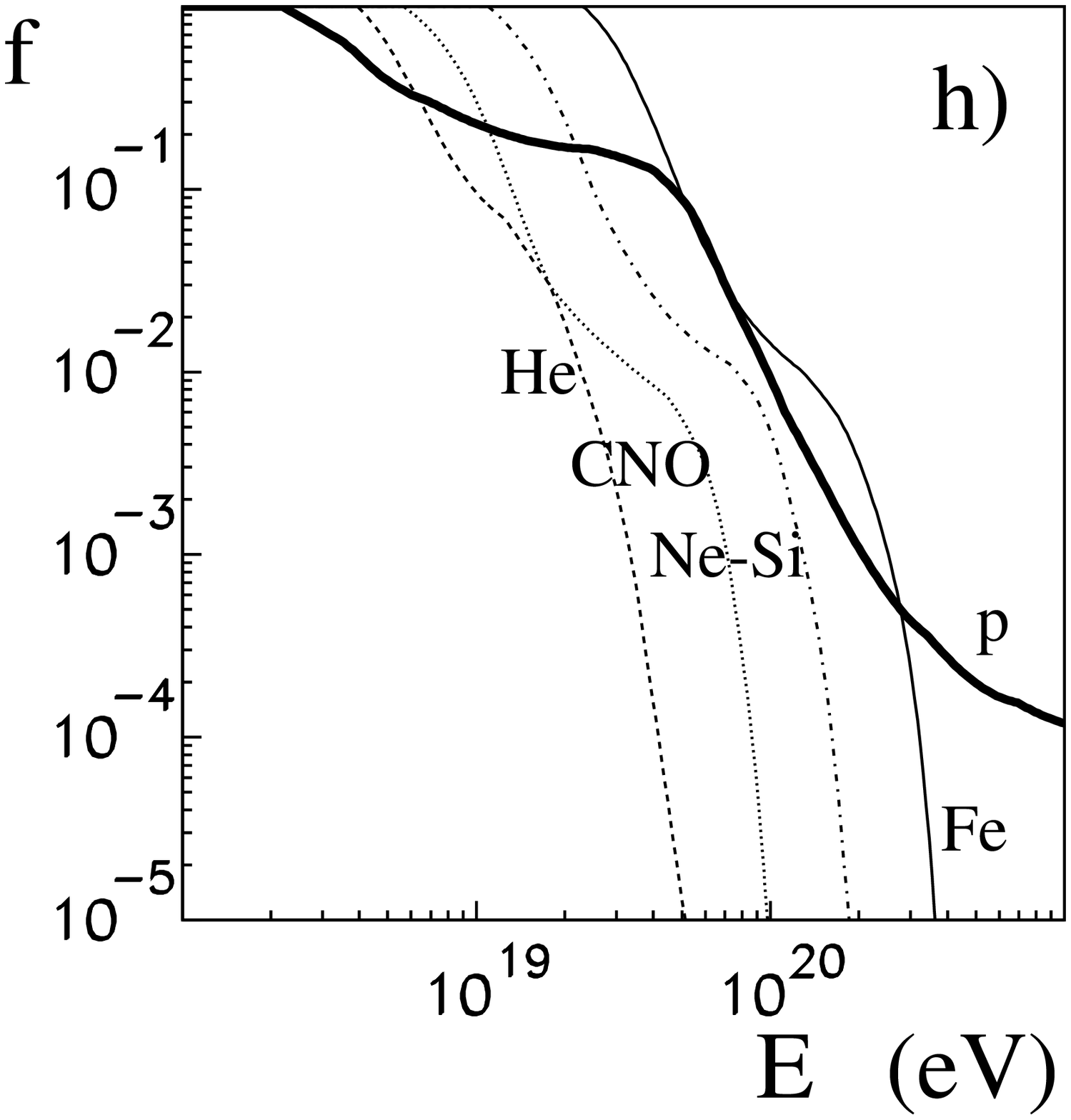}}
\caption{Cut-off factors $f$ of Eq.(\ref{feg}) calculated for different distributed CR sources: a) - uniform, b) - uniform but cut below 50 Mpc, c) - colliding galaxies, d) - Medina-Tanco, e)  - DW,
f) - CDM,  g) - cluster of galaxies and h) quasar distributions. \label{cuts} }
%\color{red} The further calculations for CDM and Medina-Tanco are in an initial state, so results are not given below. Nothing extraordinary is expected.
\end{figure}

The differences between models look huge but if we insert them to the Eq.(\ref{fnuc}) they become not so big. 
The large distances do not effect much the observed UHECR flux. Detailed results of cut-off factors are presented in Fig.\ref{cuts}.

\subsection{Mass composition results \label{composition_fit}}
Cut-off factors presented in Fig.\ref{cuts} can be applied directly to the Eq.(\ref{feg}) and used for minimization of the 
$\chi^2$ calculated for our `universal UHECR spectrum' (Fig.\ref{ave}) and $\langle X_{\rm max} \rangle$
and  rms$_{X\rm max}$ data shown in Fig.\ref{xmaxdata}. 
The discrepancy between two experiments 
concerning measured $X_{\rm max}$ distributions, but also the inconsistency of the simulation results concerning the $\langle X_{\rm max} \rangle$
and  rms$_{X\rm max}$ data, lead to the value of $\chi^2$ out of any acceptable limits, however, this does 
not mean that the analysis has no real sense. The minimum $\chi^2$ is still the measure of the `distance' between 
theory (model) and experiment (data point). The real situation means only that the statistical interpretation of $\chi^2$ is to be
 taken with care. %Because we believe that the data on spectrum, our `universal UHECR spectrum' points represents reality with small uncertainty (the flux normalization is not significant here) we can put the extra factor for the cost functions Eq.(\ref{chi}) 
%multiplying the $\chi^2$ for $\langle X_{\rm max} \rangle$ and  rms$_{X\rm max}$ by the factor of 0.1. 

Fractions of all five groups on nuclei (\p, \he, \cno, \nesi, and \fe) for both Galactic and Extragalactic
components were found together with the Galactic and Extragalactic model parameters:
\begin{itemize}
\item[-]
normalization (of the total Galactic flux)$  {\Phi_0}_{\rm G}$
\item[-]
index $\gamma_{\rm G}$
\item[-]
position of the confinement cut-off  for one of the component (\fe) $E_{\rm Gcut, \mathsf{Fe}}$, with the respectively shifted
(according to Z) values for other component fractions
\item[-]
width of the confinement cut-off $\sigma_{\rm G}$
\item[-]
normalization   (of the total Extragalactic flux)$  {\Phi_0}_{\rm EG}$
\item[-]
index $\gamma_{\rm EG}$
\end{itemize}

The strong binding of the model predictions by the `universal UHECR spectrum' related to small uncertainties at each point is complemented with the requirement of keeping the slope of the spectrum around $10^{17}$ eV to be consistent with the data below the energy of $10^{18}$ eV can be seen in figures below.
The additional simultaneous description of $\langle X_{\rm max} \rangle$ and  rms$_{X\rm max}$ data is, in principle, not possible, but we believe that there
is possible to accept, in general, the proposed model leaving the more accurate solutions to the future works with improved 
simulations which have to be used for the interpretation of the shower development data.

\begin{figure}
\centerline{
\includegraphics[width=7.3cm]{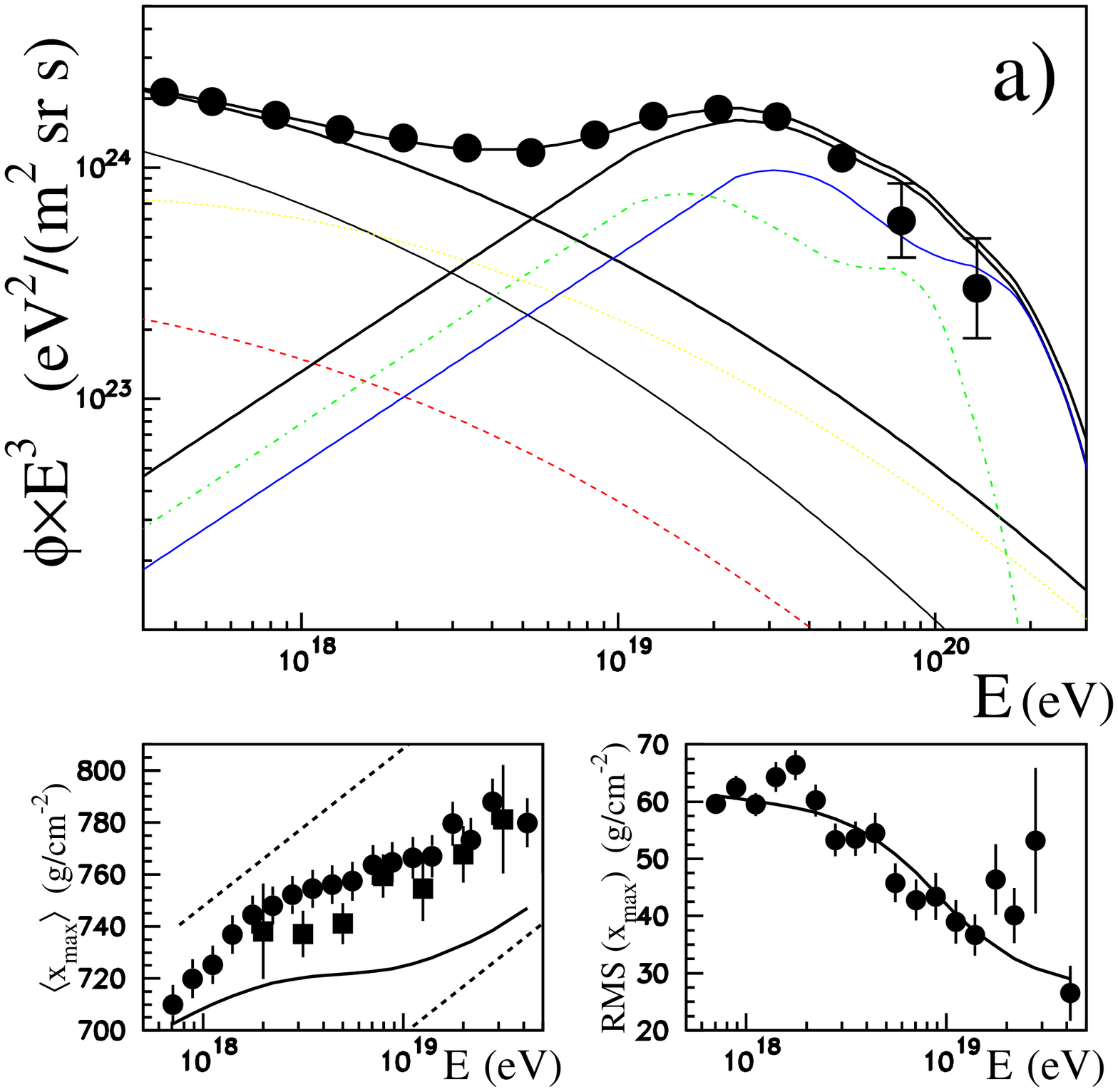}
 \includegraphics[width=7.3cm]{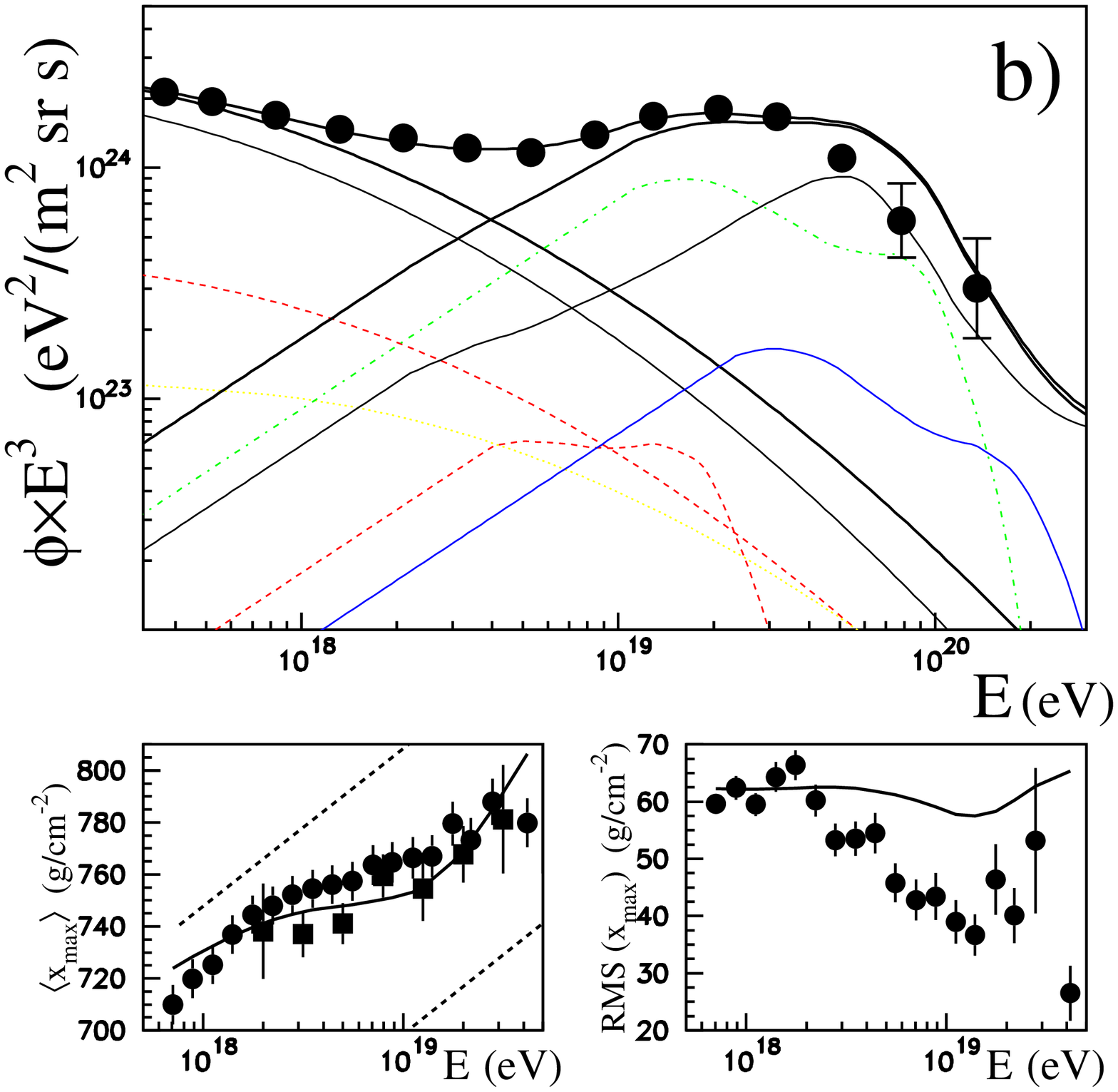}}
\caption{Model predictions for the UHECR spectrum (top) and $X_{\rm max}$ distribution data (bottom plots) with the $\langle X_{\rm max} \rangle$
 data not included in the fit (left plots) and with the rms$_{X\rm max}$   data not used for the fitting procedure (right plot).
\label{signi}}
\end{figure}

The interesting exercise is to check the quality of the fits, if we diminish the weights for the $\langle X_{\rm max} \rangle$ data (neglecting them for the moment). The satisfactory agreement can be found both for spectrum and for the  rms$_{X\rm max}$ measurements.
The $\langle X_{\rm max} \rangle$ data, of course, are then reproduces very bad (Fig.\ref{signi}a).
On the other hand the reduction of the significance of rms$_{X\rm max}$ allows us to find the good agreement  for the spectrum and 
$\langle X_{\rm max} \rangle$ data, as seen in Fig.\ref{signi}b. 

\begin{figure}
\centerline{
 \includegraphics[width=7.3cm]{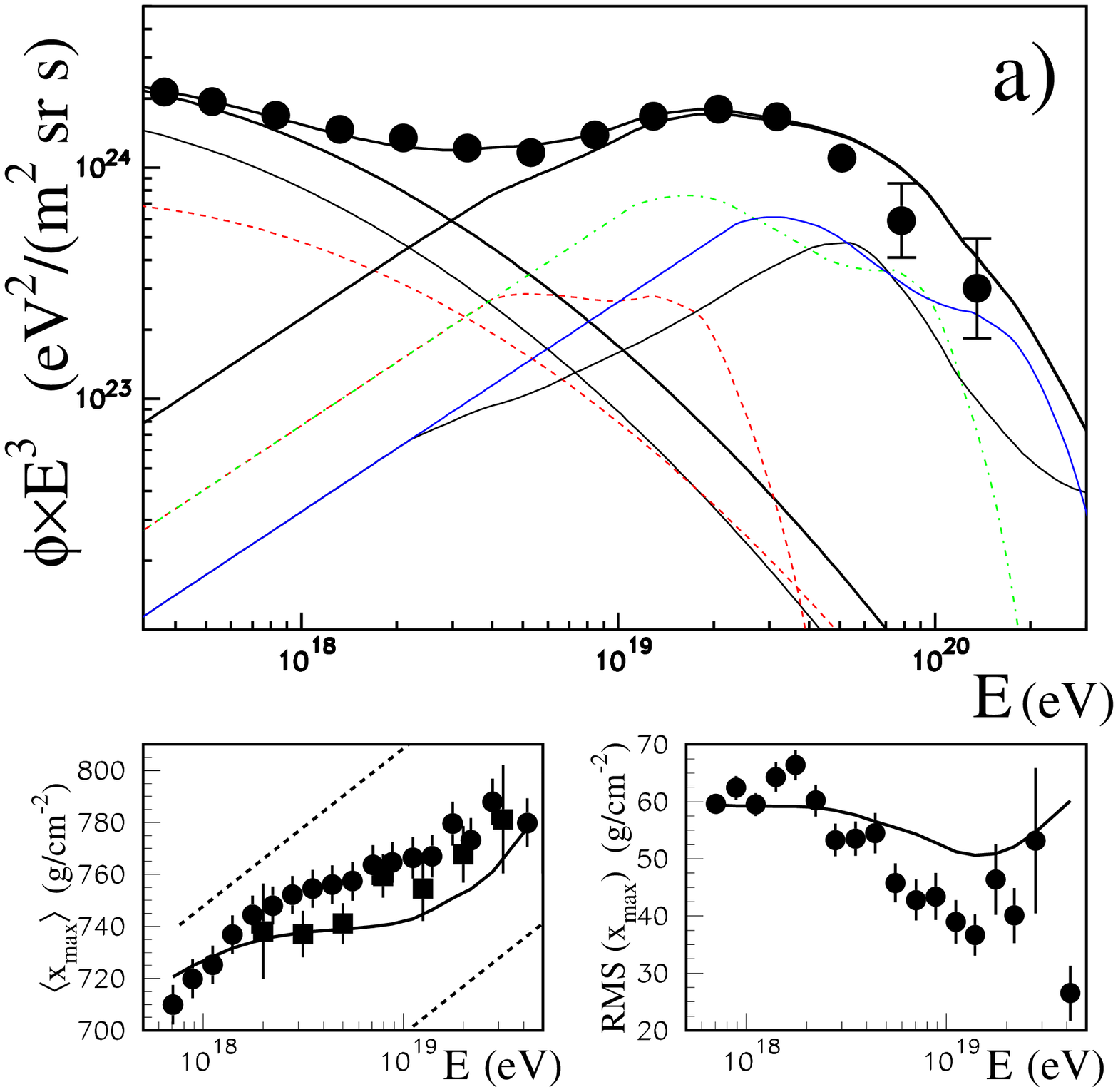}
\includegraphics[width=7.3cm]{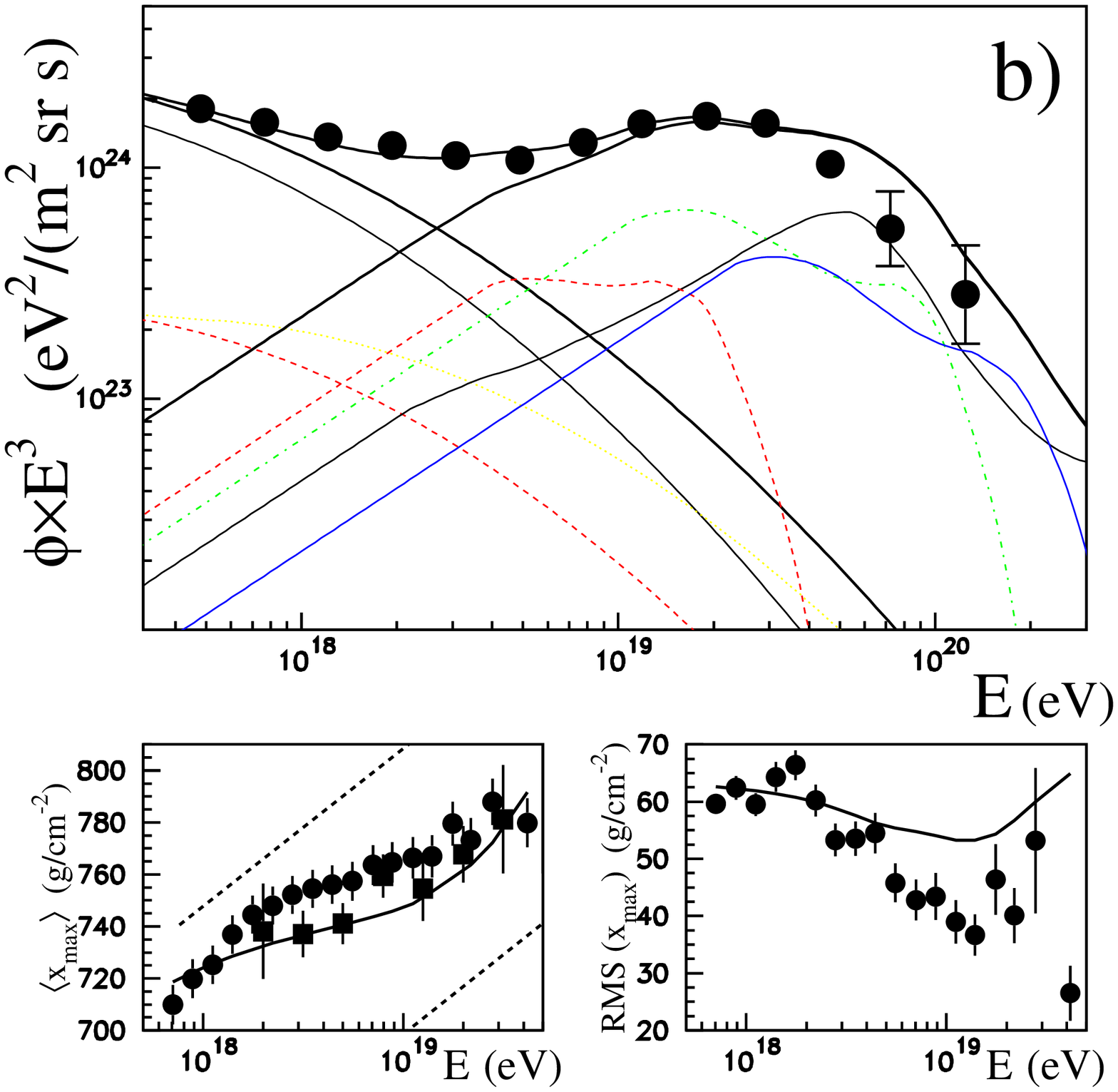}}
\caption{Model predictions calculated uniformly distributed CR sources. 
For each case the energy spectrum is presented with the components shown separately as a tiny coloured lines (\p - black solid, \he - red 
dashed, \cno - yellow dotted \nesi - dash-dotted green, \fe - blue solid), average $X_{\rm max}$ with two dashed lines representing simulation 
results for pure proton and iron predictions, and the r.m.s. of $X_{\rm max}$ with solid line obtained with PAO simulations and dashed - with TA 
simulation predictions.
Squares are the PAO measured point while circles are for TA data.  Left plots are for PAO normalization while right plots for TA.
 \label{results1}}
\end{figure}

%\textcolor[rgb]{.2,1,.2}
We would like to start our analysis from the examination of the effect of choosing different energy normalization: PAO and TA when the 'universal UHECR spectrum' is build. 
Fig.\ref{results1} shows both cases: left is for the PAO normalization, right: for the TA one. For this comparison we used 
the uniform source distribution. Results, differences due to the normalization, for other source distributions looks similar. 
The important conclusion concerning this point is a little surprising, but %the PAO and TA normalization comparison 
is looks like that the choice between PAO and TA normalization on the energy scale does not effect much 
the composition found. The differences are quite small and statistically not significant.
For the PAO normalization the extragalactic component consists of about 0.5 of light fractions (\p \ and \he) and the same amount of heavies (\nesi \ and \fe) while
for TA the ratio of light to heavy is 0.6 to 0.4.

%{\color{red}
%there are too many plots in Fig.\ref{results1}. I think that two pairs are quite enough. For example: uniform and the one with the 50 Mpc cut-off. Rest will be deleted. I keep them here for completness.}

\begin{figure}
\centerline{
 \includegraphics[width=6.cm]{fig10a.eps}
 \includegraphics[width=6.cm]{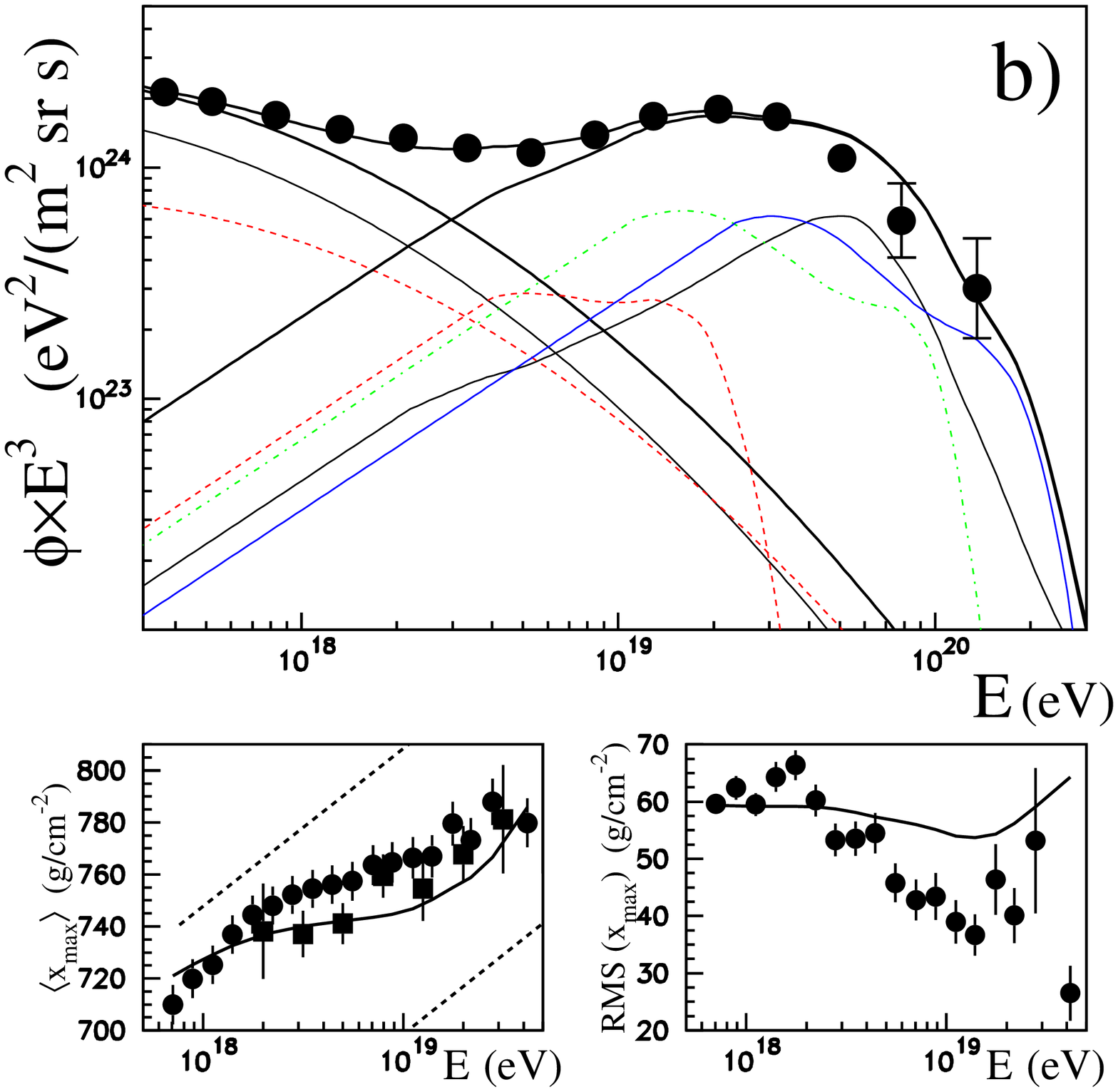}}
\centerline{
 \includegraphics[width=6.cm]{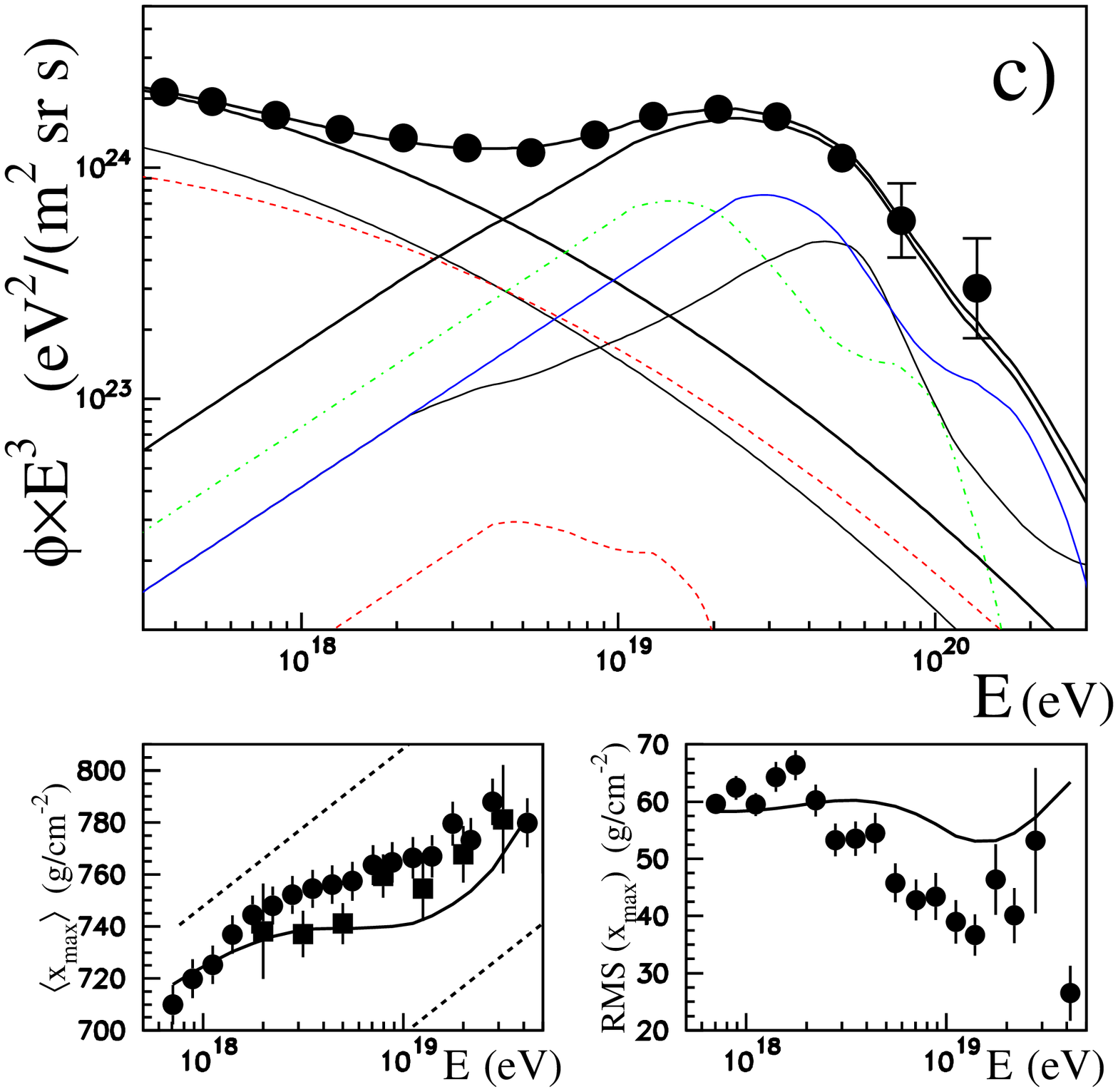}
 \includegraphics[width=6.cm]{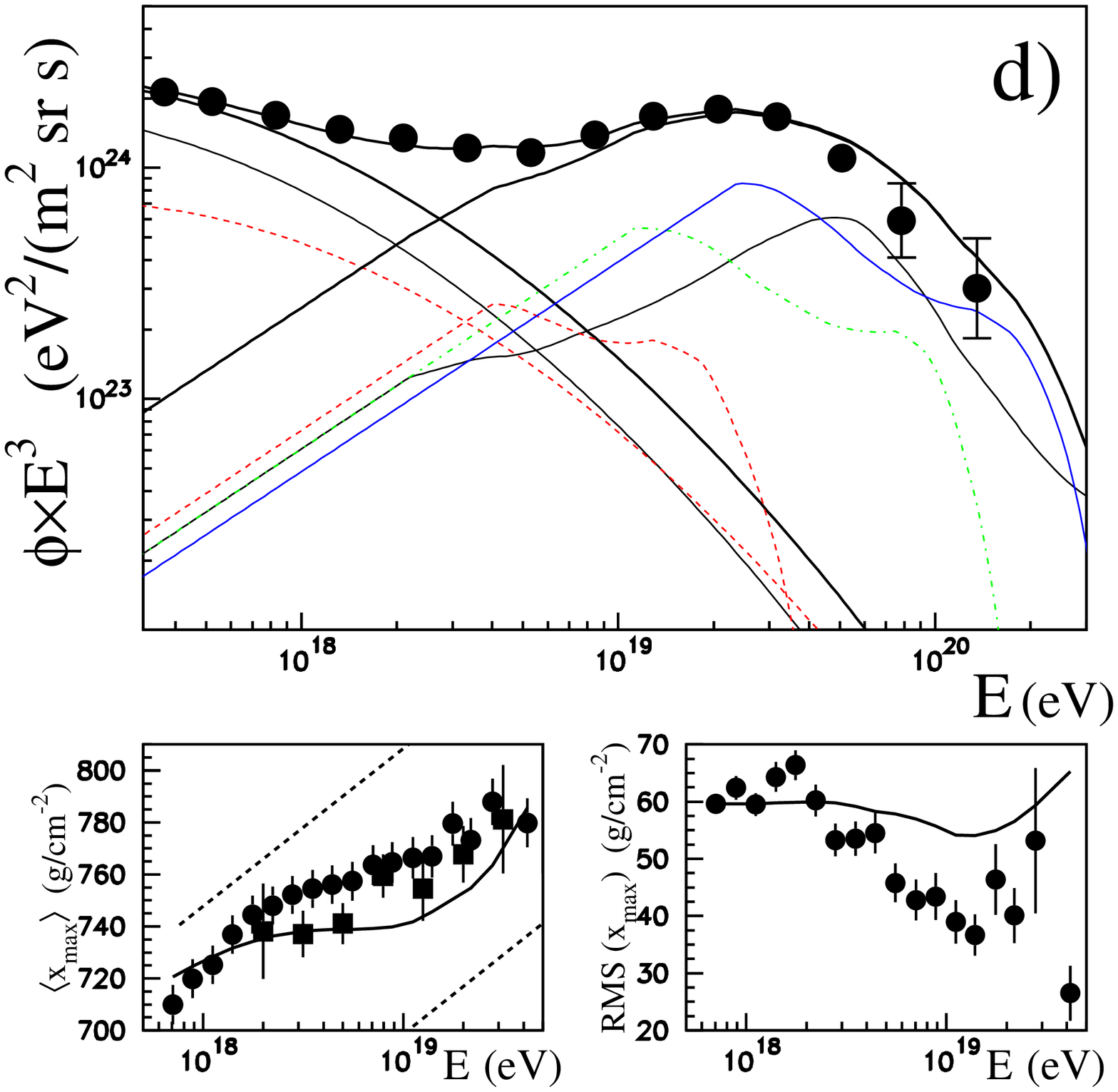}}
\centerline{
 \includegraphics[width=6.cm]{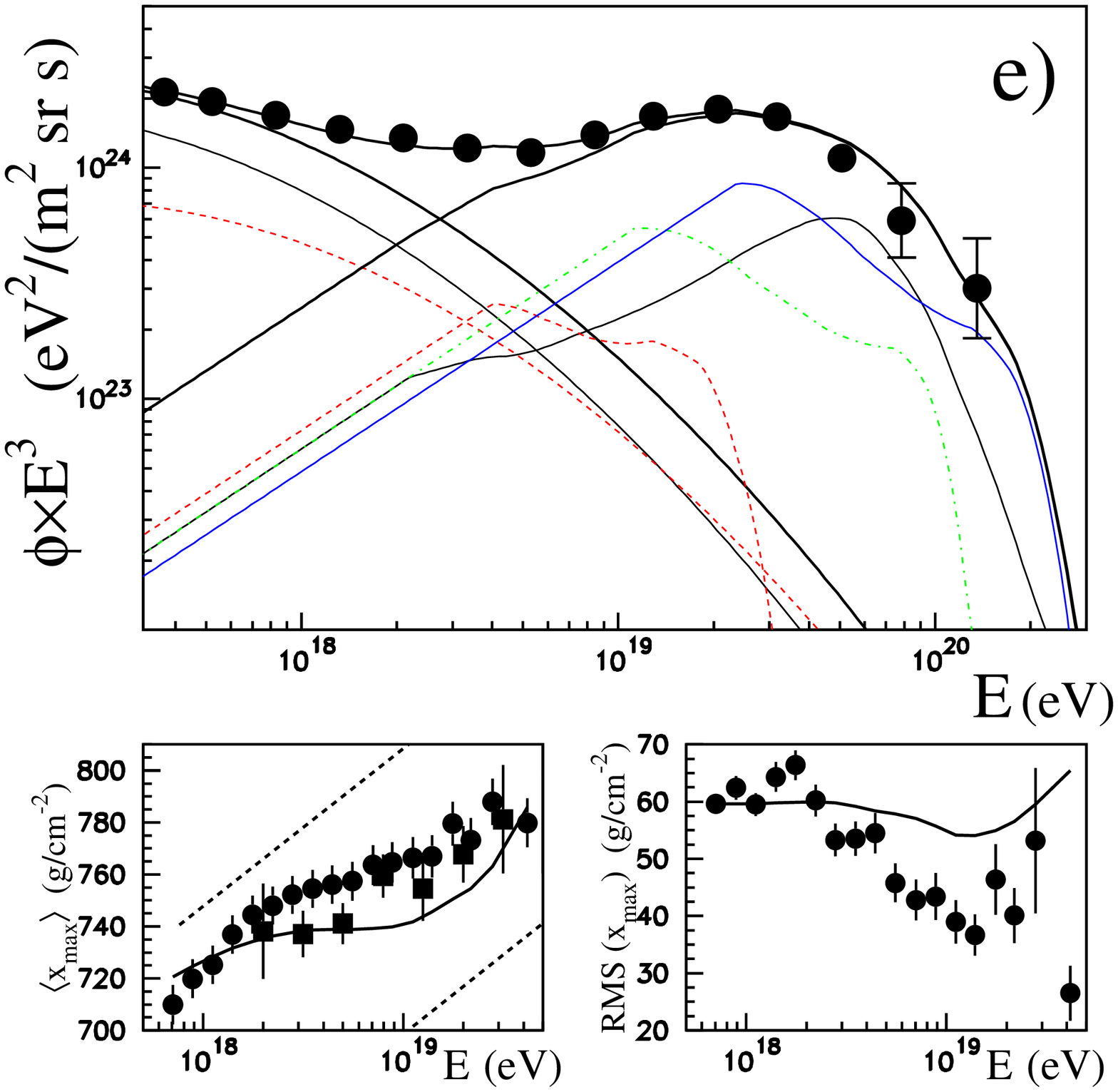}
 \includegraphics[width=6.cm]{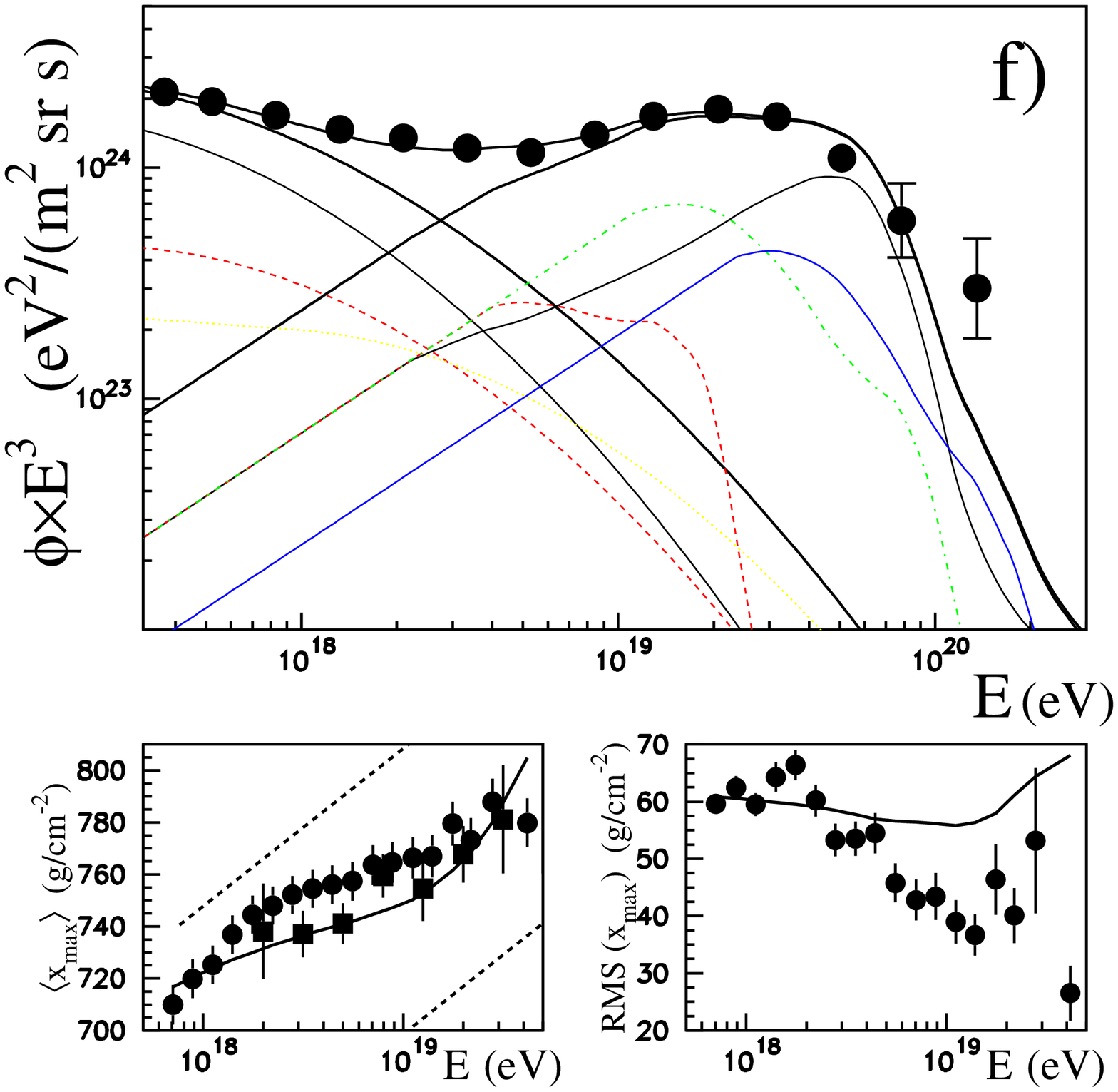}}
\centerline{
 \includegraphics[width=6.cm]{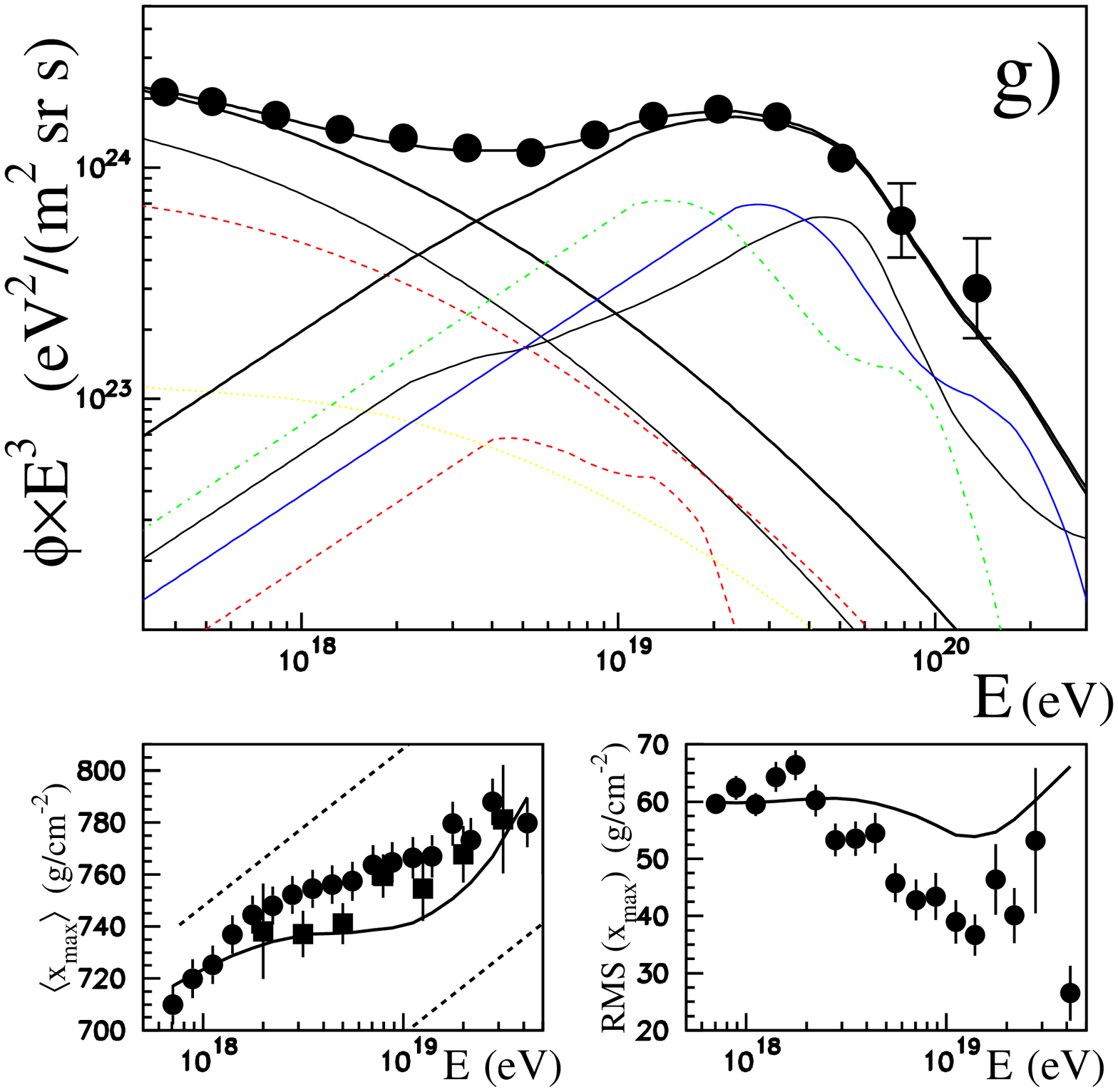}
 \includegraphics[width=6.cm]{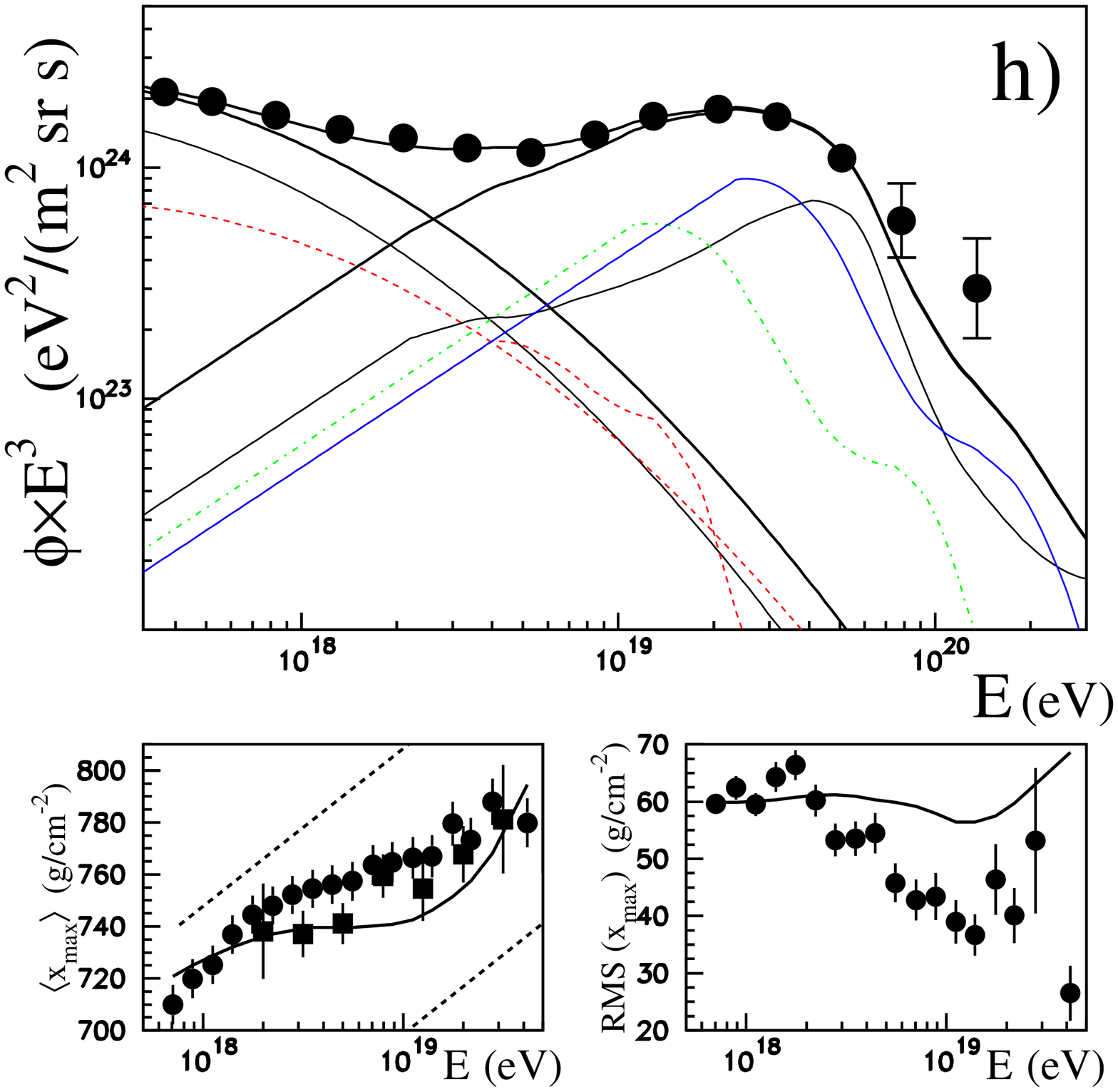}}
 \caption{Results calculated for different distributed CR sources:  uniform (as shown in Fig.\ref{results1} - left) - a), cut at 50 Mpc b), colliding galaxies c),  Medina-Tanco proposition d), DW e), CMD f)  
cluster of galaxies g), and  quasar h) distributions with PAO normalization.
 \label{results3}}
\end{figure}

Keeping this in mind we present in Fig.\ref{results3} results of adjusting the source composition for all analyzed 
source distribution models obtained for PAO energy normalized `universal UHECR spectrum'.

In Fig.\ref{results3} the separate nuclei components were plotted. We can see some differences there. Respective fractions of each group of nuclei in 
Galactic and Extragalactic UHECR flux are listed in the Table~\ref{fractions}.  Fractions obtained with PAO normalization of the energy are given in the 
first row, while for TA energy normalization in the second row for each analyzed distribution.

\begin{table}
\caption{\label{fractions}Fractions of the Galactic and Extragalactic mass components for different source distributions and two normalizations of UHECR universal spectrum.}
\begin{ruledtabular}
\begin{tabular}{r|ccccc|ccccc}
 &\multicolumn{5}{c|}{Galactic}&\multicolumn{5}{c}{Extragalactic}\\
\hline
source distribution& \p & \he& \cno& \nesi& \fe& \p & \he& \cno& \nesi& \fe \\
\hline
%uniform {\small (PAO norm.)}\  & \input{../composition/cut-off-3-uniform/fr.txt}
%{\small (TA norm.)}\ & \input{../composition/cut-off-4-uniform/fr.txt}
%cut below 50 Mpc\ & \input{../composition/cut-off-3-0to50/fr.txt}
% & \input{../composition/cut-off-4-0to50/fr.txt}
%colliding galaxies\ & \input{../composition/cut-off-3-coll/fr.txt}
% & \input{../composition/cut-off-4-coll/fr.txt}
%DW% \cite{dudarewicz}
 %                       \ & \input{../composition/cut-off-3-duda/fr.txt}
 %& \input{../composition/cut-off-4-duda/fr.txt}
%galaxy clusters\ & \input{../composition/cut-off-3-clusg/fr.txt}
 %& \input{../composition/cut-off-4-clusg/fr.txt}
%quasars\ & \input{../composition/cut-off-3-quasar2/fr.txt}
% & \input{../composition/cut-off-4-quasar2/fr.txt}
%\input{rys/fraction.tex}
uniform {\footnotesize (PAO norm.)}\  & 
   0.70 &    0.30 &    0.00 &    0.00 &    0.00 &    0.15 &    0.35 &    0.00 &    0.35 &    0.15 \\
\ {\footnotesize (TA norm.)}\  & 
   0.80 &    0.10 &    0.10 &    0.00 &    0.00 &    0.20 &    0.40 &    0.00 &    0.30 &    0.10 \\
cut below 50 Mpc \  & 
   0.70 &    0.30 &    0.00 &    0.00 &    0.00 &    0.20 &    0.35 &    0.00 &    0.30 &    0.15 \\
 & 
   0.80 &    0.10 &    0.10 &    0.00 &    0.00 &    0.20 &    0.40 &    0.00 &    0.30 &    0.10 \\
colliding galaxies \cite{0954-3899-22-12-013}\  & 
   0.60 &    0.40 &    0.00 &    0.00 &    0.00 &    0.25 &    0.05 &    0.00 &    0.45 &    0.25 \\
 & 
   0.75 &    0.05 &    0.00 &    0.20 &    0.00 &    0.30 &    0.35 &    0.00 &    0.25 &    0.10 \\
M-T  \cite{1538-4357-510-2-L91}\  &  
   0.70 &    0.30 &    0.00 &    0.00 &    0.00 &    0.25 &    0.30 &    0.00 &    0.25 &    0.20 \\
 & 
   0.70 &    0.30 &    0.00 &    0.00 &    0.00 &    0.25 &    0.35 &    0.00 &    0.25 &    0.15 \\
DW \cite{dudarewicz}\  & 
   0.70 &    0.30 &    0.00 &    0.00 &    0.00 &    0.25 &    0.30 &    0.00 &    0.25 &    0.20 \\
 & 
   0.70 &    0.30 &    0.00 &    0.00 &    0.00 &    0.25 &    0.25 &    0.00 &    0.35 &    0.15 \\
CDM \cite{wwmexico} \  & 
   0.70 &    0.20 &    0.10 &    0.00 &    0.00 &    0.30 &    0.30 &    0.00 &    0.30 &    0.10 \\
 & 
   0.70 &    0.30 &    0.00 &    0.00 &    0.00 &    0.15 &    0.20 &    0.00 &    0.45 &    0.20 \\
galaxy clusters   \cite{Bahcall:1988ch} \  & 
   0.65 &    0.30 &    0.05 &    0.00 &    0.00 &    0.30 &    0.10 &    0.00 &    0.40 &    0.20 \\
 & 
   0.60 &    0.40 &    0.00 &    0.00 &    0.00 &    0.20 &    0.00 &    0.00 &    0.50 &    0.30 \\
quasars \  & 
   0.70 &    0.30 &    0.00 &    0.00 &    0.00 &    0.35 &    0.20 &    0.00 &    0.25 &    0.20 \\
 & 
   0.70 &    0.30 &    0.00 &    0.00 &    0.00 &    0.30 &    0.20 &    0.00 &    0.30 &    0.20 \\\end{tabular}
\end{ruledtabular}
\end{table}

\begin{figure}
\centerline{
 \includegraphics[width=8.5cm]{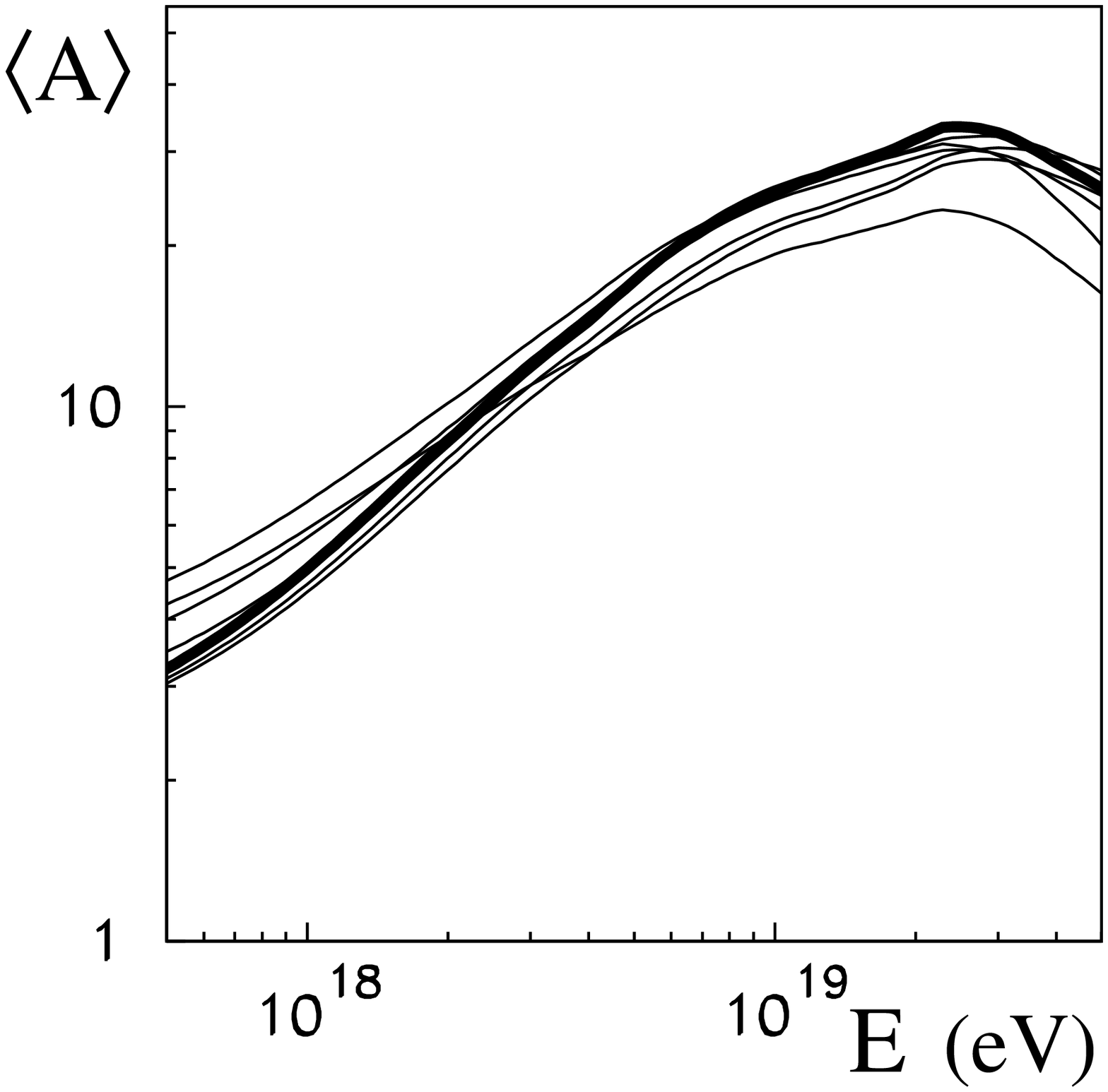}}
 \caption{Average UHECR particle mass  as a function of particle energy. Each line represents different source distribution model analysed in this work. The differences are not big. The bold line is the result of the model Dudarewicz \& Wolfendale \cite{dudarewicz}.
 \label{mas}}
\end{figure}

Fractions shown in the Table \ref{fractions} and individual element spectra shapes allows us to show the average mass of the UHECR as a function of the registered particle energy. Results are given in Fig.~\ref{mas}. The possible systematics related to the CR source distribution is space for the energies in the ankle region is of order of 20\%.

%------------------------------------------------------------------------------------------------
\section {Discussion \label{discusion}}
In the paper \cite{Szabelski2002125} we have combined spectra obtained by five greatest experiment of these times Hi-Res, Fly's Eye, Haverah Park, Yakutsk and AGASA using the idea of the universality of the feature called `the ankle'. We were able to describe the
`universal spectrum' obtained as the sum of Galactic and Extragalactic fluxes with significant fraction of heavy nuclei at the CR sources. We then put forward the question {\it At what particle energy do extragalactic cosmic rays start to predominate?} in Ref. \cite{0954-3899-31-3-005} answering positively in the same paper. Further analysis confirmed this \cite{0954-3899-34-9-003, wwmexico} and with the first PAO results in Refs. \cite{Wibig:2007pf, astra-7-275-2011}.

In the present paper which takes into account the recent data from PAO and TA experiments the `universal UHECR spectrum' obtained using both measured spectra and additionally the  $X_{\rm max}$ results we confirmed the idea of exchanging the
Galactic to Extragalactic CR flux at the `ankle'. The confirmation is made stronger by the composition data analysis which 
shows the significant fraction of heavy element above $10^{19}$ eV, above the `ankle'.

The `universal spectrum' obtained in this work allows us to examine the distribution of Extragalactic sources  of UHECR. Some models are more advantageous than others. The important contribution to the spectrum shape is the density of the EG sources in the vicinity (tens of Mpc) from Earth. The lack of galaxy clusters seen in \cite{1992ApJ...388....9T, dudarewicz} shows the
right directions to search for the UHECR sources, however the statistics of extremely high energy events is still too small to make the more definite statement. 

We can compare our results with the main results of the PAO measurement of CR spectrum and shower development concerning the Extragalactic component shape and composition
\cite{1475-7516-2017-04-038}. Using the uniform source distribution model (with the standard cosmological evolution) they
conclude that the EG UHECR flux consists of heavy nuclei ($\sim$80\% of \cno \ and rest of \nesi \ group) with hard and very hard spectrum, $\gamma=2.04\pm 0.01$  ($\gamma=0.96\pm 0.1$ for the `main minimum' ).
We believe that their `second minimum' (in the $\chi^2$ search) is the real one. The value of the 'main minimum' is reported as 'unexpected' even by the Authors: Boncioli, di Matteo and Grillo from the Pierre Auger Collaboration \cite{BONCIOLI2016139} in their paper title {\it ``Surprises from extragalactic propagation of UHECRs"}).
Such hard spectrum index is probably the effect of using the particular phenomenological model of EG component with the extra free parameter describing the rigidity dependent particle energy cut-off introduced alongside the known GZK and photodisintegration limits which work with no doubts at the very end of the cosmic ray spectrum. 
The very good agreement of the measured by PAO flux, and  $X_{\rm max}$ data with their model predictions is reported, and it is no surprise.
The hard injection spectrum is obtained also in \cite{PhysRevD.92.123001}
where Unger, Farrar and Anchordoqui discussed their quite flexible model of the `ankle'.

The TA experimental team, following the HiRes Collaboration conclusions \cite{PhysRevLett.104.161101}, reported rather opposite results concerning the UHECR mass composition. 
In Ref.~\cite{Abbasi2016131} the TA Collaboration confirmed the HiRes and HiRes/MIA results indicate of $\sim$ 50\%
of protons at $10^{17}$ eV, increasing to $\sim90$\% for energies above
10$^{18}$ eV with the $+20$\%$-40$\% uncertainty. This wide `error band' accommodate almost all possible scenarios.
The significant fraction of protons in the Galactic component is supported by the anisotropy results 
\cite{2016MNRAS.460.3479T}.
In Refs. \cite{Tinyakov201429, Fukushima:2015bza} TA
spectra and  $X_{\rm max}$ data used as average values $\langle X_{\rm max}\rangle$ but also as the whole set of distributions is narrow energy bins favour clearly the pure proton cosmic ray flux at the whole energy range $10^{18}$ --$10^{20}$ eV.

The analysis similar to the one presented above but only for the PAO results concerning the spectrum and $X_{\rm max}$ data 
has been shown in Ref.\cite{PhysRevD.92.063011} by Taylor, Ahlers and Hooper, in Ref. \cite{Hooper2010151} by Hooper and Taylor
and by Taylor in Refs.\cite{taylorthesis, Taylor201448}. One of the conclusions \cite{Taylor201448} is the significant fraction of  medium light (\cno \ and \nesi) nuclei with the hard ($\gamma <2$)  injection spectrum. They studied also an interesting aspect, the effect of magnetic field strength (and structure) on the spectra (and anisotropy).
In  Ref. \cite{PhysRevD.84.105007} Taylor, Ahlers and Aharonian the importance of nearby UHECR sources is analyzed. Their result of the of absence of close ($R<20$--$50$ Mpc) sources is to some extent confirmed by our calculations (the slightly better fit with the Dudarewicz \& Wolfendale \cite{dudarewicz} model). 

 Globus, Allard and Parizot in \cite{PhysRevD.92.021302} present the perfect agreement of the model they propose in the energy region of our interest with the PAO data on spectrum and  $X_{\rm max}$. The model assumes the Galactic and Extragalactic UHECR components with mixed composition dominated by protons below $5 \times 10^{18}$ eV then dropping to 10\% at  $3 \times 10^{19}$ eV with introduction in the transition region of the dominated here \cno \ nuclei group. There are different spectra (slopes) for EG protons, and all other EG components: \he, \cno \ and heavier nuclei up to \fe. With different source hardnesses and different cut-offs for each nuclei group the agreement in the `ankle' region and above is not very surprising.
 
\section{Conclusions}

We have examined the UHECR spectra measured by two big experiments The Pierre Auger Observatory and The Telescope Array Project. Both delivered numerous collections of events, the Extensive Air Showers initiated by ultra-high energy cosmic rays observed by arrays of ground detectors and by fluorescent light telescopes. The sophisticated technique used for their registration was complemented by complex procedures of the raw data correction and the corrected data evaluation to eventually form the reported UHECR particle energy spectra. The spectra measured in southern hemisphere by PAO and in northern by TA are displaced slightly both in the energy scale and on the intensity. This issue is known for years and is still standing open. We have tried to
go through it and we obtained two solutions for the `world average' UHECR spectrum with PAO and TA normalization, respectively. Some part of the discrepancy was reduced by adjusting possible, but a priori unknown difference in the individual shower energy reconstruction accuracy. It was estimated by both experimental group to be of order of 20\%. After applying this procedure the difference in the energy normalization in both cases was found not to be as big as initially reported. 

We then used both spectra for the further analysis of the mass composition. 
The study of UHECR energy spectrum alone, as it is mentioned already about 10 years ago in Ref.~\cite{1475-7516-2008-10-033} can not give the definite, conclusive
answer if the composition is heavy of light, or very light. Both proposition are acceptable. 
To find the answer one has to use the data on the shower longitudinal development. We have take them in the form of
average depth of the shower maximum $\langle X_{\rm max}\rangle$ and its dispersion rms$_{X\rm max}$.

We assumed that the UHECR flux is the sum of the the Galactic and Extragalactic components given by the simple (simplest?)
form of power-laws at the sources with two indexes for Galactic of order of 3.1 and for Extragalactic of 2.1. Modified by the rigidity dependent confinement for the Galactic components and by the GZK or photodisintegration processes for the Extragalactic one.

Both components were assumed to be, in principle, mixtures of five group of nuclei from single nucleons, to iron nuclei. The four free parameters of each component (eight all together) were to be found comparing the model predictions with the PAO and TA data.
 
Our findings can be summarized as: 
\begin{itemize}
\item[-]
the discrepancy of the energy scales between PAO and TA experiments  does not effect
substantially results of the mass composition of UHECR,
\item[-]
the `universal UHECR spectrum' can be reproduced for any reasonable source model distribution rather well,
\item[-]
the light composition of the Galactic component of UHECR is mostly caused by the $\langle X_{\rm max} \rangle$ data
of the PAO; it has to be remembered that this result is based strongly on the (PAO) shower simulations,
\item[-]
the composition of extragalactic CR is heavier, it consists of about half of the \fe \ and \nesi \ group nuclei, but there is non 
negligible fraction of light elements, mostly protons and  \he, what is caused by the rms$_{X\rm max}$ data,
\item[-]
the medium \cno \ group of nuclei for both components, G and EG, is not needed at all to explain the data,
\item[-]
the possible systematics of the source composition (the average mass) related to the CR source distribution for the energies in the ankle region is estimated to be of order of 20\%,
\item[-]
there is no possibility to answer definitely to the question of the distribution of the UHECR sources in space, but:
\begin{itemize}
\item[i)]
some recommendations can be given to the distribution without CR sources in the vicinity of the Earth, e.g., Dudarewicz \& Wolfendale \cite{dudarewicz} or `colliding galaxies',
\item[ii)]
models assuming the strong cosmological evolution as `galaxy clusters' and even more the `quasar' model underestimates the flux above 10$^{20}$ eV,
\item[iii)]
the uniform CR sources distribution, and the proposition of Medina-Tanco \cite{1538-4357-510-2-L91} slightly overestimate UHECR flux at around 10$^{20}$ eV,
\end{itemize}
\item[-] 
the common shape (the same slope) of the UHECR at the sources of the EG component leads to the definite difficulty with the explanation of the deep penetrating of UHECR in the atmosphere. It can suggest the needs of the revision of the ultra-high energy interaction model.
\end{itemize}

The last item is, maybe, the less conclusive, but in our opinion very important. The presented above short review of some ideas and quantitative analysis shows that further progress concerning the origin and nature of extremely high energy cosmic rays 
needs new insight to the very high energy interactions of nuclei which is not a simply, or not so simple, but still a superposition of many individual proton-proton multiparticle production processes. \cite{PhysRevLett.117.192001, PhysRevD.91.032003}.
%,doi:10.1093/astrogeo/atw209}

%{\color{red} but the calculations are still going on and some solutions may change}
%\nocite{*}
\bibliography{WW2017_03}% 

\end{document}